\documentclass[bibyear]{aa}
\usepackage{graphicx,gensymb}
\usepackage{txfonts}
\begin{document}

   \title{A color-period diagram for the open cluster M\,48 (NGC\,2548), \\and its rotational age\thanks{Based on data obtained with the STELLA robotic telescopes in Tenerife, an AIP facility jointly operated by AIP and IAC; this paper presents results for the STELLA Open Cluster Survey (SOCS); Appendix A is only available in electronic form via http://www.edpsciences.org; the cluster photometry table is only available in electronic form at the CDS via anonymous ftp to cdsarc.u-strasbg.fr (130.79.128.5) or via http://cdsweb.u-strasbg.fr/cgi-bin/qcat?J/A+A/}
}

%   \subtitle{I. Subtitle}

   \author{Sydney A.~Barnes\inst{1,2}, 
           Joerg~Weingrill\inst{1},
           Thomas~Granzer\inst{1},
           Federico~Spada\inst{1},
           \and
           Klaus G.~Strassmeier\inst{1}
%         \and
%         Other Authors\inst{2}\fnmsep\thanks{Just to show the usage
%          of the elements in the author field}
          }
   \titlerunning{Color-period diagram for M\,48}
   \authorrunning{Barnes, Weingrill, Granzer, Spada \& Strassmeier}

   \institute{Leibniz Institute for Astrophysics Potsdam (AIP), 
                 An der Sternwarte 16, 14482 Potsdam, \and
              Space Science Institute, 
                 4750 Walnut Street, Boulder, CO 80301, USA\\
          \email{sbarnes,jweingrill,tgranzer,fspada,kstrassmeier@aip.de}
%              \email{sbarnes@aip.de}
%         \and
%             University of Alexandria, Department of Geography, ...\\
%             \email{c.ptolemy@hipparch.uheaven.space}
%             \thanks{The university of heaven temporarily does not
%                     accept e-mails}
             }

   \date{Received Month date, year; accepted Month date, year}
%   \date{Received September 15, 1996; accepted March 16, 1997}

% \abstract{}{}{}{}{} 
% 5 {} token are mandatory
 
  \abstract
 {Rotation periods are increasingly being used to derive ages for cool single field stars. Such ages are based on an empirical understanding of how cool stars spin down, acquired by constructing color-period diagrams (CPDs) for a series of open clusters.
 Our main aims here are to construct a CPD for M\,48, to compare this with other clusters of similar age to check for consistency, and to derive a rotational age for M\,48 using gyrochronology.
 We monitored M\,48 photometrically for over 2\,months with AIP's STELLA\,I 1.2\,m telescope and the WiFSIP 4K imager in Tenerife. Light curves with 3\,mmag precision for bright (V$\sim$14\,mag) stars were produced and then analysed to provide rotation periods. A cluster CPD has then been constructed.
 We report 62 rotation periods for cool stars in M\,48. The CPD displays a clear slow/I-sequence of rotating stars, similar to those seen in the 625\,Myr-old Hyades and 590\,Myr-old Praesepe clusters, and below both, confirming that M\,48 is younger. A similar comparison with the 250\,Myr-old M\,34 cluster shows that M\,48 is older and does not possess any fast/C-sequence G or early K stars like those in M\,34, although relatively fast rotators do seem to be present among the late-K and M stars. A more detailed comparison of the CPD with rotational evolution models shows that the cluster stars have a mean age of 450\,Myr, and its (rotating) stars can be individually dated to $\pm 117$\,Myr (26\%). Much of this uncertainty stems from intrinsic astrophysical spread in initial periods, and almost all stars are consistent with a single age of $450$\,Myr. The gyro-age of M\,48 as a whole is 450$\pm$50\,Myr, in agreement with the previously determined isochrone age of 400$\pm$100\,Myr.}


  \maketitle
%
%________________________________________________________________

\section{Introduction}

The study of stellar rotation, both among field stars and in open clusters, 
was synonymous with $v\,sin\,i$ measurements for several decades. See Kraft
(1970), and references therein, for a review of developments until 1970.
Good starting points for developments into the 1980s and 1990s, especially 
those relating to open clusters,  are
Stauffer \& Hartmann (1987), Soderblom et al. (1993), and Queloz et al. (1998).
More recently the emphasis has shifted to photometric rotation
period measurements, partly because they avoid the $sin\,i$ ambiguity inherent 
to spectroscopic measurements.

Such measurements began with the pioneering work of 
Van Leeuwen, Alphenaar \& Meys (1987)\footnote{See also Van Leeuwen \& Alphenaar (1982).}, who measured rotation periods photometrically for 11 cool stars in the Pleiades open cluster, and were soon followed by the remarkable Hyades 
rotation-period measurements\footnote{No older cluster was measured successfully until Meibom et al. (2011b) studied the 1\,Gyr-old cluster NGC\,6811 using the Kepler Space Telescope.} of Radick et al. (1987). The Mt. Wilson sample of stars,
 with periods measured from variability in chromospheric emission 
(Baliunas et al. 1996) and rooted in earlier spectroscopic work, including the 
discovery of stellar chromospheric activity cycles (Wilson 1978), is also 
noteworthy.

A large body of subsequent work by the astronomical community 
[e.g., Bouvier et al. 1993 (T\,Tauri stars), Barnes et al. 1999 (IC\,2602), 
 Irwin et al. 2007 (NGC\,2516), Meibom et al. 2009 (M35), Hartman et al. 2010 
 (Pleiades)] 
has shown that rotation period measurements constitute a distinct new probe of 
stellar evolution that provides both similar and new information, as compared 
with `classical' methods, such as isochrone fitting of color-magnitude diagrams 
(CMDs), from which it is steadily becoming independent. 
Such independence is valuable because young clusters contain few (or sometimes
even no) bona fide giant members, making isochrone fitting particularly 
challenging.

Indeed, it is now recognized that color-period diagrams (CPDs) of open clusters 
are similar to CMDs, and provide a useful complementary means of characterizing 
open clusters, particularly in terms of ranking them by age 
(e.g., Barnes 2003, Meibom et al. 2015). 
Making the reasonable assumption that stars in both open clusters and the
field spin down in similar ways because spindown is governed by processes 
internal to stars allows field star ages and those of any accompanying planets 
to be derived using gyrochronology (e.g., Barnes 2007), a valuable ability in 
the Kepler (and soon PLATO) era.

%{\bf Sequences: }\\
Another notable aspect of open cluster CPDs is that they 
often display sequences of rotating stars (Barnes 2003).
Clusters like the Hyades (Radick et al. 1987; Delorme et al. 2011) and 
NGC\,6811 (Meibom et al. 2011b) display a clear slow or I-sequence 
(consisting of relatively slowly-rotating stars whose periods increase steadily 
as cool star spectral types change from F- to G- to K-type). 
These stars have converged onto this sequence over the several-100\,Myr ages of 
these clusters. 
Zero-age main sequence clusters, such as IC\,2391 (Patten \& Simon 1996) and 
IC\,2602 (Barnes et al. 1999), display only weak evidence for such a sequence. 
(However, the paucity of available stars could play a role in the 
difficulty of recognizing it in such cases.)

Certain young clusters, such as M\,35 (Meibom et al. 2009), and the
Pleiades (Hartman et al. 2010), display evident fast/C-sequences of
rotating stars (consisting of many G-, K-, and M-type stars with rotation 
periods, $P \lesssim$\,1\,d), 
in addition to the slow/I-sequence that characterizes older clusters.
This fast sequence appears to dissipate rapidly on a timescale that depends 
on stellar mass, because these stars spin down to populate the cluster's slow 
sequence instead.
This is still a subject of active research. For recent ideas, see 
Matt et al. (2015), Gallet \& Bouvier (2015), Brown (2014), 
Epstein \& Pinsonneault (2014), and references therein.
Earlier ideas concerning slow- and fast-rotating stars (with varying emphasis
on the mass dependence of rotation suggested by rotational sequences) can be 
found in, e.g., MacGregor \& Brenner (1991), Chaboyer et al. (1995), Collier-Cameron et al. (1995), Barnes \& Sofia (1996), Bouvier et al. (1997), and Sills et al. (2000).

%{\bf Role of M\,48: }\\
A significant part of the difficulty in understanding the morphological
changes that occur between the Pleiades-type ($\lesssim$200\,Myr) clusters 
and the Hyades-type ($\sim$600\,Myr) clusters is the lack of appropriate 
published rotation-period observations for clusters of intermediate age.
The only one available to date is the M\,37 cluster (Hartman et al. 2009),
whose rotational age has been claimed to be younger than its 540\,Myr 
isochrone age.
With an isochrone age of 400\,Myr (Balaguer-Nunez et al. 2005), 
M\,48 lies squarely in this intermediate age range and could help in 
elucidating the transitional rotational behavior.
%{\bf Justification for this work: }\\
It is therefore desirable to construct a reliable CPD to characterize the 
M\,48 cluster and to explore its properties empirically in the context of 
other well-studied clusters. 
In particular, a CPD would allow a comparison between the age determined from 
rotation and the age determined from classical isochrone fitting.

More generally, such studies also increase our basic knowledge 
(e.g., photometry and membership) of the often unknown lower main sequence 
populations of these clusters, a difficult task in the pre-CCD era. 
We have studied the open cluster \object{M 48} (\object{NGC 2548})
in this context. An additional context is provided by the STELLA Open Cluster
Survey, which aims to provide rotation periods for a series of open clusters.
For a related study of the IC\,4756 cluster, see Strassmeier et al. (2015).

%{\bf Prior literature: }\\
Despite being a Messier (1781) object\footnote{The cluster apparently lies $2.5 \degree$ south of Messier's position, and was identified with NGC\,2548 by Oswalt Thomas in 1934.}, 
M\,48 ($\alpha_{2000}$ = 08 13 43, $\delta_{2000}$ = --05 45 00) 
has not been the subject of many prior studies, a fact that may be related
to its being located in the southern sky.
%Ebbinghausen (1939) 
Ebbighausen (1939) performed a proper motion study of the upper main
sequence (B and A spectral types) and giants of the cluster within a 
$15^{\prime}$ radius, and identified 74 of these stars as probable members.
%Wu et al. (2002)
Another dedicated cluster study was not published for more than 60 years,
until Wu et al. (2002) performed another proper motion study of a  
$1.6 \times 1.6 \degree$ region around the cluster and identified 165
stars as probable cluster members. This particular paper did not provide 
any photometric information.
%Balaguer-Nunez, L., Jordi, C \& Galadi-Enriquez, D. (2005)
However, a related study by Balaguer-Nunez, Jordi \& Galadi-Enriquez (2005;
hereafter BJG05) has revealed that the Wu et al. (2002) astrometric cluster 
members were brighter than $V \approx 13$. BJG05 focused on providing 
multicolor Stromgren photometry for a cluster-centered 
$34^{\prime} \times 34^{\prime}$ region to a depth of $V \approx 22$, and using 
this photometry they constructed a candidate member list to a depth of $V = 18$.
They also (re-)determined the cluster's basic parameters, including the age, 
400${\pm}$100\,Myr. % (${\rm log}\,t = 8.6 \pm 0.1$).

%Rider et al. (2004)
Almost contemporaneously, Rider et al. (2004) provided photometry in the 
Sloan filter system for the cluster region to a depth of $g^{\prime}_0 = 16$ 
and found that a 400\,Myr isochrone matched their photometry reasonably 
well.
%Wu et al. (2006)
Finally, the most recent work on this cluster by Wu et al. (2006) 
provided 13-band photometry in a specialized (BATC) filter set over a 
$58 ^{\prime} \times 58 ^{\prime}$ field, and identified 323 stars as 
(SED-based) photometric candidate cluster members. An agreement level of 
80\% was claimed between this membership criterion and earlier 
proper motion studies. 
% Pesch (1961)
Curiously, no photometry in Johnson colors is currently available for this
cluster beyond the bright stars studied photoelectrically by Pesch (1961) 
and the ($V, I$) study of Sharma et al. (2006), the latter restricted to stars 
within an $8^{\prime}$ radius of the cluster center. 
% Our emphasis
The study presented here builds on these prior ones with an emphasis on the 
rotational properties of the cluster's stars.

%{\bf Motivations for this paper: }\\
%We provide CCD-based Johnson B,V photometry to a depth of V$\sim$20,\\ 
%rotation periods, \\
%a CPD,\\ 
%and various other items in this paper.

%{\bf Layout of the paper: }\\
The rest of this paper is organized as follows.
The observations are discussed in Section\,2.
We present the cluster color-magnitude diagram in Johnson B and V
colors to a depth of V$\sim$20 in Section\,3.
The variability analysis is presented in Section\,4, 
leading to the construction of the cluster color-period diagram, 
followed by relevant comparisons.
Finally, the conclusions are presented in Section\,5.

%__________________________________________________________________

\section{The Observations}

%{\bf STELLA: }\\
The M\,48 open cluster was observed with the WiFSIP 4K CCD imager mounted on
the AIP's STELLA\,I robotic 1.2\,m telescope, located at the IAC in Tenerife, 
Spain (Longitude: $16\degree 30^{\prime} 35^{\prime \prime}$ West, 
Latitude: $28\degree 18^{\prime} 00^{\prime \prime}$).
The STELLA robotic observatory (also containing a 1.2m spectroscopic telescope,
STELLA\,II) opens and closes automatically every usable night, guided by
measurements of a number of weather and meteorological parameters. 
During this interval, it observes a set of targets that are chosen by a 
scheduling program on the basis of user-defined priorities. 
Details about the facility and its operation may be found in 
Strassmeier et al. (2004), Granzer (2004), and 
Strassmeier, Granzer \& Weber (2010).

%{\bf Time-series observations: }\\
The cluster was observed nightly (as allowed by weather) over a two-month 
baseline from 4\,March\,2014 to 7\,May\,2014. 
A $44^{\prime} \times 44^{\prime}$ field, consisting of a $2 \times 2$ mosaic of 
$22^{\prime}$ CCD fields (NW, NE, SW, SE) centered on the cluster, was 
monitored. 
Our field is significantly larger than the region covered by 
BJG05, as can be seen in Fig.\,1.
Having this large a field was fortuitous, because cluster stars and rotators 
extend across the entire region monitored. In fact, the cluster very likely 
extends significantly beyond even our study field\footnote{For instance, the rotators we have identified below extend to the edges of the observed field.}.
Generally, the telescope cycled sequentially through the four subfields.
The time-series exposures were acquired in the Johnson V\,band with
exposure times of 30\,s (short) and 300\,s (long). 
The number of visits (either exposure time) per night per field ranged from 
zero to 7, as permitted by other scheduled programs, weather, and telescope 
performance.

Over the observing period, we acquired a total of 1481 short- and long-exposure
frames for the four fields. Of these, we (conservatively) discarded 992 because 
of tracking errors, lunar proximity, bad seeing, or conditions deemed too far 
from photometric for our purposes, leaving us with 489 high-quality visits to 
the four cluster fields (average = 122/field) for the time series observations. 
% NW: 137 frames, NE: 108 frames, SE: 109 frames, SW: 135 frames
Small differences in the numbers of images for each sub-field arise from 
variations in observing conditions, and because our observing program did not 
modify individual field observing priorities after the acquisition of 
incomplete nightly observing cycles.

% Fig.1 Sky coverage
   \begin{figure}
   \centering
   \includegraphics[width=100mm]{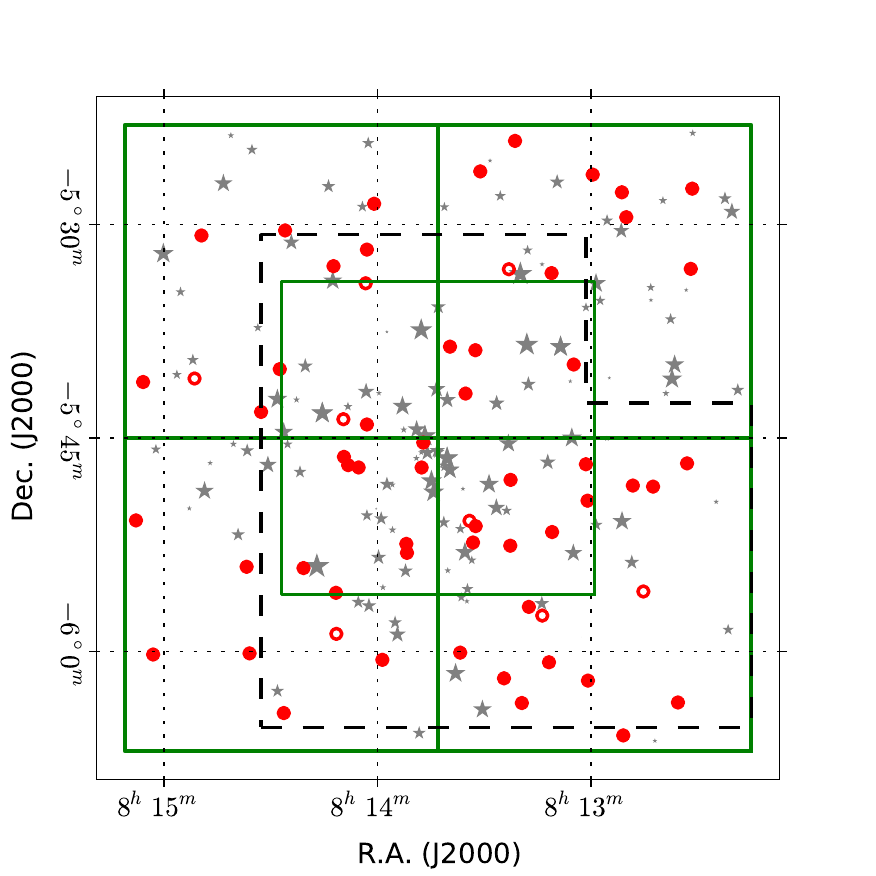}
      \caption{On-sky areal coverage of our photometry (a $2 \times 2$ CCD mosaic covering a $44^{\prime} \times 44^{\prime}$ region and an additional central field) in the region of the M\,48 cluster is displayed with solid gray lines (green online). The dashed lines indicate the (smaller) region covered by BJG05. Stars brighter than $V = 12$ are marked, with symbol sizes scaled inversely with their V magnitudes. The locations of the 62 identified rotating stars (discussed later) are also indicated with circular symbols.              }
         \label{fig1}
   \end{figure}

%{\bf Color-Mag Obs: }\\
In addition, sets of Johnson B and V exposures were obtained for each of
the cluster sub-fields on multiple nights, together with standard star
observations of Landolt (2009) standard fields. 
This allowed us to obtain colors for the CPD, and of course, 
to construct a CMD for the cluster in Johnson colors.
However, because the number of standard stars observed was small, we decided
to place our final photometry on the photoelectric system of Pesch (1961).
We also specifically acquired photometry for a separate field bore-sighted on
the cluster center, to verify and ensure that the $2 \times 2$ mosaic of 
cluster sub-fields are placed on a common photometric system. 

%{\bf 2 Paras from Thomas about moving data from the telescope to Joerg: }\\
As part of the robotic observations, bias and flat field frames are obtained 
before and after science observations. Two bracketing master bias frames, each 
consisting of 25 individual frames, are used for bias subtraction on the 
science frames. No dark current subtraction is performed as the dark current is 
below 1e$^{-}$/h at the nominal operation temperature. With the limited time 
available for twilight sky flats, it is impossible to flat field all of the 
21 filters of WiFSIP on a single night. Instead, each twilight phase is used 
to calibrate 2-3 filters with 10 individual flat exposures each, grouped 
in two five-exposure sequences at opposite derotator settings to level-out 
first order illumination gradients remaining in twilight sky flats for imaging 
instruments with large fields of view (Chromey \& Hasselbacher, 1996).
The master 
flat field for individual science frames is then constructed by averaging at 
least ten such flat field blocks, but allowing for up to 100 blocks as long as 
a maximum time difference between flat block and science frame of less than 
$\pm$25 days is not exceeded. The small variations in image scale across the 
field of view lead to each pixel receiving light from differing solid angles, 
in turn leading to differing light levels even on perfectly flat-illuminated 
fields. The average master flat field is corrected for this geometrical effect 
before usage on the science image. 

For multi-amplifier readout modes like ours, crosstalk between the 
amplifiers can lead to ghost signals on corresponding pixel positions. Although 
it is a minor effect of a few tenths of a per cent, raw images are 
corrected for crosstalk. Subtle gain variations between the different 
amplifiers  are compensated for by adjusting the amplifier gain to allow for a 
steady transition in illumination level across the amplifier read-out edges. 
Again, these variations are a few tenths of a per cent, 
compared to the nominal gain factors of $1.568$ and $1.587$, for the two 
amplifiers used; however, the precision desirable for our time-series analysis 
makes such compensation advisable.
Before performing the final photometry, the astrometric world-coordinate 
solution is obtained using a modified version of WCSTools (Mink 2002).
The final solution follows the FITS conventions (Calabretta \& Greisen 2002)
for a zenithal polynomial projection (ZPN) to third degree, with the remaining 
root-mean square (rms) error in the position on the order of $0.2$ arcsec.

\subsection{Photometry}

%{\bf From Joerg: }\\
%{\bf The first step is the calibration of the images.} 
Photometry for each frame was performed with SExtractor 
(Bertin \& Arnouts, 1996) using the {\sc isocorr} aperture. 
This choice was preferable to others such as {\sc fixed aperture} or
{\sc isoauto} because it worked better than those choices in this work, 
principally because our point spread function (PSF) is sometimes distorted by 
tracking errors and/or less-than-perfect seeing conditions.
Using an adaptive aperture preserves the most frames for photometry, 
with minimal impact on the photometric quality.

For each field, the frame with the highest ratio of matched to identified 
sources in the PPMXL catalog
%({\bf presumably refers to Thomas' section. Clarify!})
was selected as the reference frame. 
This selection was also verified manually and showed that only photometric 
nights were chosen. 
The reference frames for all five fields (2$\times$2 mosaic + central
field) ended up being selected from only a few photometric nights 
(28\,Feb\,2014 to 7\,Mar\,2014). The particular fields standardized on each 
night are listed in Table\,1.

\begin{table}
\caption{Dates and numbers of secondary standard stars} % title of Table
\label{table:1}      % is used to refer this table in the text
\centering                          % used for centering table
\begin{tabular}{l c c}        % centered columns (4 columns)
\hline\hline
Field & Date & Number of stars\\
\hline
M 48 BVI NE & 2014-02-28 & 184 \\
M 48 BVI NE & 2014-03-01 & 184 \\
M 48 BVI NW & 2014-03-03 & 192 \\
M 48 BVI C  & 2014-03-06 & 211 \\
M 48 BVI SE & 2014-03-05 & 209 \\
M 48 BVI SE & 2014-03-02 & 208 \\
M 48 BVI SW & 2014-03-07 & 205 \\
\hline
\end{tabular}
\end{table}

Up to five frames having the smallest offsets with respect to the reference 
frame for stars in the range between 10 to 16 magnitudes were then selected. 
%The reference frame is naturally included in this sample by having a zero offset.
These offsets calculated from the mean differences between the stars in the 
reference frame and the cross-identified ones in the selected frames were 
typically less than 4 millimag, with an rms below 1 millimag. 
The mean magnitude for each star was calculated from the five selected frames.
The nightly offset of the reference frame to the Landolt standard fields was 
finally added to the derived magnitudes.
Only stars within a 0.15" matching radius were identified as cross-matched.
The whole procedure was performed separately for the B and V frames. 

%{\bf This was moved here from below: }\\
%We have also obtained Landolt (2009) standards on every useful night to 
%enable us to assess the photometricity, or any deviation therefrom.
%This photometry is essentially on an identical system, with an average 
%systematic difference of no more than {\bf $0.00X$\,mag} in $V$, and 
%{\bf $0.00X$\,mag} in $B-V$ on the nights of {\bf XX-YY\,Apr2014}, 
%our best standard nights.

%\newpage
 
\section{Color-magnitude diagram}

%{\bf REWRITE: }\\
B and V frames from the best photometric nights were manually selected for 
the CMD. For these frames the successive magnitude 
determinations for the best-exposed stars were repeatable at the 3\,mmag level.
We decided simply to place the frames on the (photoelectric) 
photometric standard system of Pesch (1961).

We matched 36 stars from Pesch (1961) listed on Simbad and Webda with our
dataset. Webda lists 37 stars from Pesch and 14 stars from Oja (1976, 
priv. comm.). Since the Oja sample included fainter stars, and proved to be 
of similar photometric quality, we included them in our calibration sample.
Four stars out of the initially 40 cross-identified ones were omitted for
calibration due to large differences in $V$ magnitude or $B-V$ color (TYC
4859-28-1, HD 68779, BD-05 2451, BD-05 2452). This might arise from
coordinate mismatching or bad photometry. Eventually 30 stars from Pesch and 
6 stars from Oja were retained for calibration. 
None of our calibration stars showed saturated pixels in the PSF.
The $V$ magnitude of the calibration stars covered a range from 8.184\,mag to 
14.272\,mag and $0.001 < B-V < 1.441$ in color.
For these 36 stars we achieved an rms of 0.032\,mag in 
$V_{\rm STELLA}$-$V_{\rm Pesch/Oja}$
and 0.020 mag in $(B-V)_{\rm STELLA}$ -- $(B-V)_{\rm Pesch/Oja}$. 

%{\bf Body: }\\
The CMD in Johnson $B-V$ color for the entire 
$44^{\prime} \times 44^{\prime}$ survey area of our study is displayed in the
upper panel of Fig.\,2.
The cluster's main sequence is obvious, and is reasonably distinguishable 
visually from the background stars in this direction of the Galaxy.
The cluster sequence begins somewhat brighter than $V = 10$, which is where 
the A-type cluster stars are located, and extends diagonally downward to 
$V \sim 19$, where the cluster's M-type stars become indistinguishable 
from the field population. 
A binary sequence, while undoubtedly present, is not visually prominent.
A hint of a white dwarf sequence is visible.
%The cluster giants are likely overexposed in our photometry. 
%We therefore recommend using prior sources referenced above for 
%photometry of the brightest cluster members. For instance, Pesch (1961) has 
%provided photoelectric $UBV$ photometry for 37 of the brightest stars.
Our scientific interests center on the fainter (V > 13) cool FGK stars. 
%which are therefore optimally exposed. 
Photometric information for all stars and cross-identifications with BJG05 are 
provided in an accompanying table\footnote{This table is only available in 
electronic form at the CDS via anonymous ftp to cdsarc.u-strasbg.fr 
(130.79.128.5) or via http://cdsweb.u-strasbg.fr/cgi-bin/qcat?J/A+A/
}.

% Fig.2 Color-magnitude diagrams for the cluster
   \begin{figure}
%   \centering
% 2 upper
    \includegraphics[angle=-90,trim=0cm 3cm 0cm 0cm,clip=true,width=105mm]{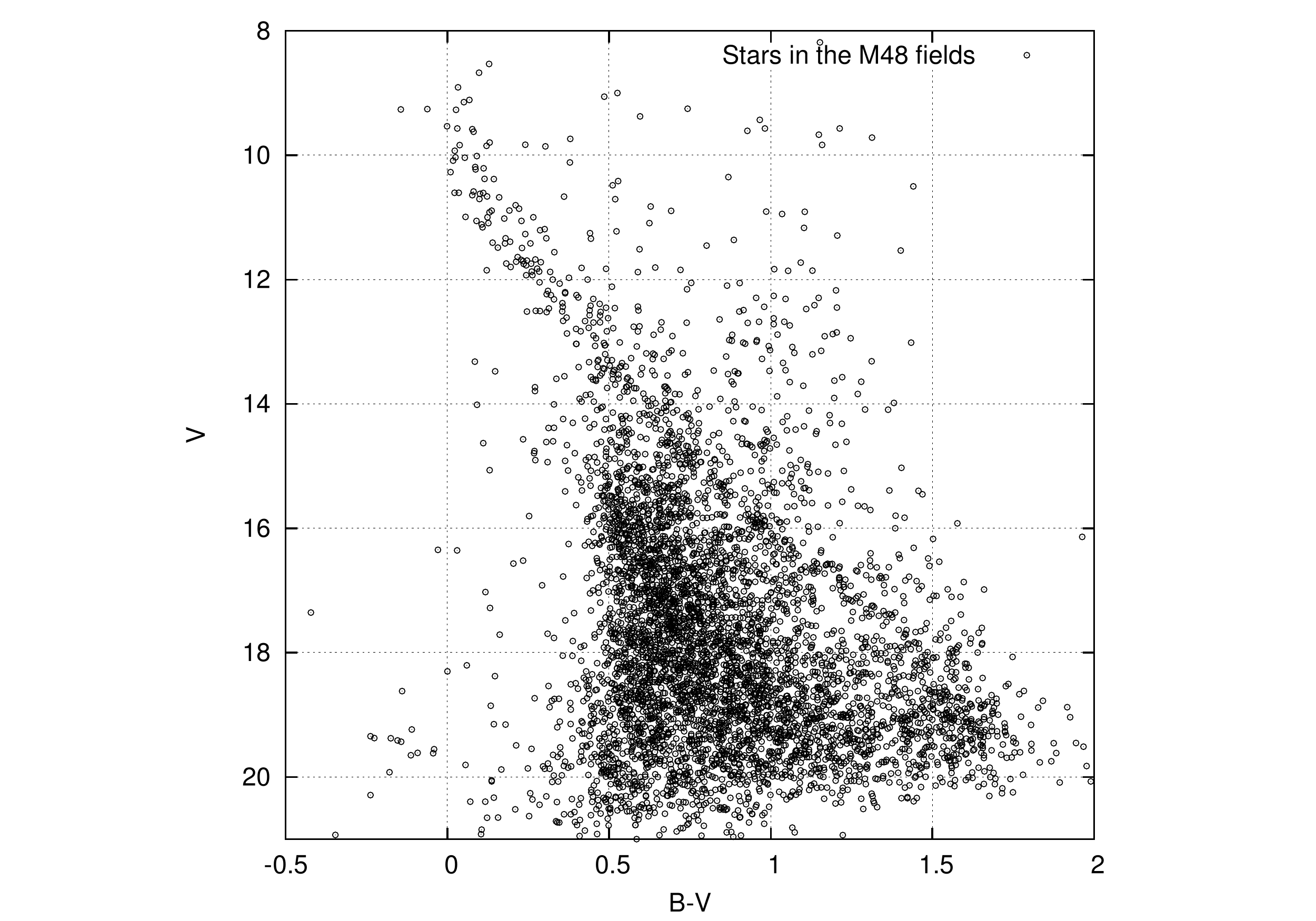}
% 2 lower
    \includegraphics[angle=-90,trim=0cm 3cm 0cm 0cm,clip=true,width=105mm]{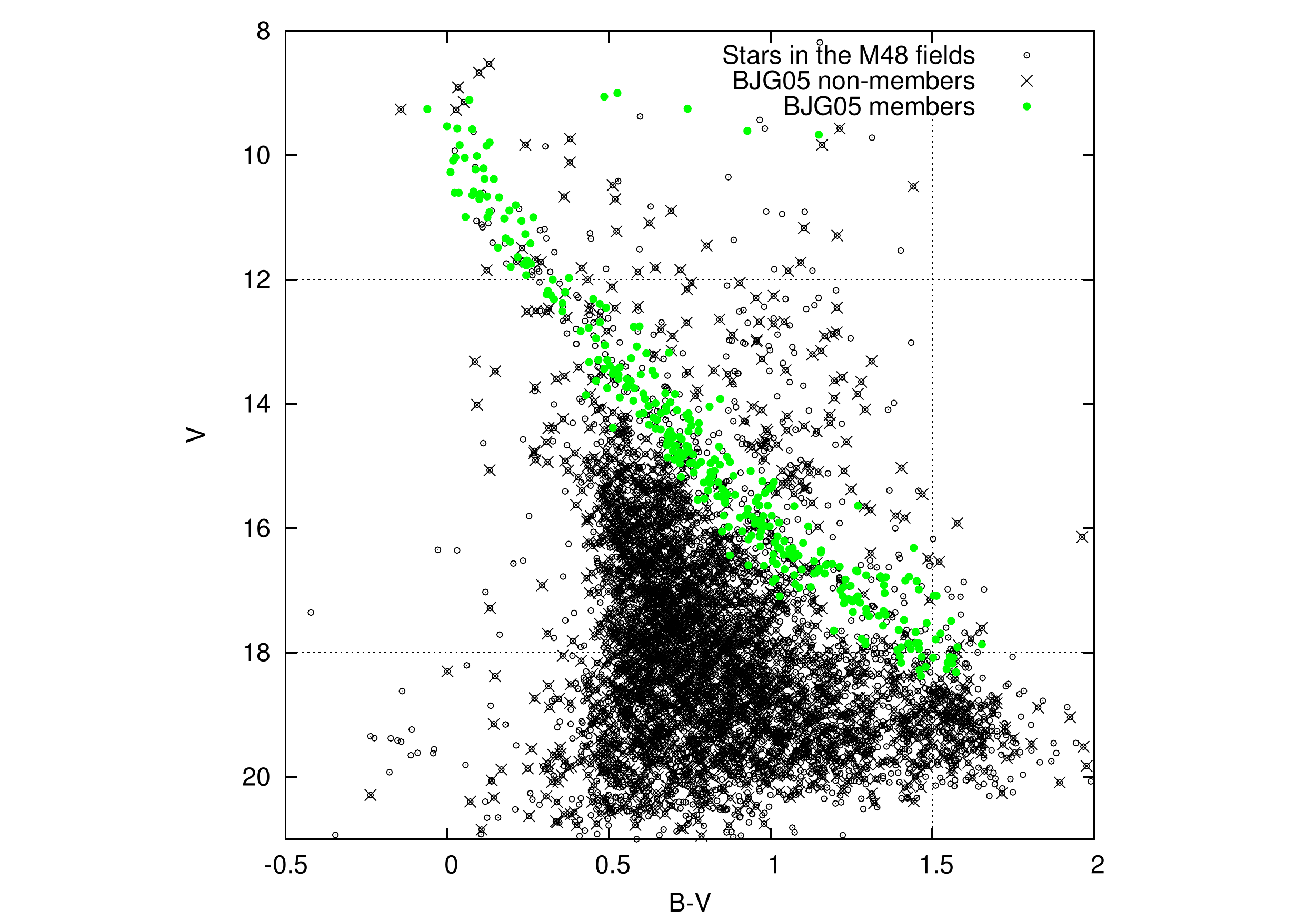}
      \caption{{\it Upper Panel:} Color-magnitude diagram (CMD) for the $44^{\prime} \times 44^{\prime}$ M\,48 cluster region of our study in Johnson $B-V$ color. A cluster sequence, beginning brighter than $V=10$ and extending diagonally downward to $V \sim 19$, is seen clearly against the field population. 
%A binary sequence is not visually prominent.
{\it Lower panel:} Large filled gray circles (green online) in the M\,48 CMD indicate cluster members identified by BJG05, based on astrometry brighter than $V = 13$, and multi-color Stromgren photometry for fainter stars. Most of these BJG05 members are clearly on the cluster sequence in our photometry. Stars proposed by BJG05 to be non-members are additionally indicated with crosses, while the remaining empty circles have no BJG05 membership information. 
%{\bf Make 2-column}
      }
 \label{fig2}
\end{figure}

%{\bf Membership info: }\\
It is desirable to understand which of these stars have been determined to be 
cluster members by prior studies, and to use this information where possible.
The deepest of these studies is that by BJG05.
Beginning with the Wu et al. (2006) astrometric members, which run out at
a depth of $V \sim 13$, they extended the candidate member list downward to
$V \sim 18$ using multi-color Stromgren photometry to select cluster members.
We have cross-identified the cluster members selected by Balaguer-Nunez et 
al. (2005) against our photometry, and have marked these stars in the 
color-magnitude diagram displayed in the lower panel of Fig.\,2. 
(As an aside, we note that BJG05 defined the cluster sequence using an 
empirical ZAMS constructed by Crawford (1975) and succeeding authors as 
referenced in BJG05.)
We observe that the vast majority of the members identified by 
BJG05 indeed lie on the cluster's sequence in our 
($B, V$) photometry. It therefore appears that the BJG05 selection is an 
inclusive, rather than an exclusive one.

While we are fortunate that this prior membership information from 
BJG05 is available for a significant fraction of the 
stars of interest in our study, our study area is somewhat larger than theirs 
(see Fig.\,1), so it will not be a surprise that we are able to propose 
additional candidate cluster members, some of them even with determined periods.
Accordingly, a number of our rotational candidate cluster members (see Table\,2
below) are classified as ``--'', indicating that they do not have a BJG05
designation. However, these stars are both on the cluster's photometric
sequence in our ($B, V$) color-magnitude diagram, and have measured rotation 
periods consistent with cluster membership.
(In Table\,2, M = BJG05 member, and N = BJG05 non-member.)
Ultimately, the BJG05 cluster member selection and ours 
are both photometric below $V = 13$, the primary region of interest for this 
work, and therefore neither can be considered definitive in this region. 

%{\bf Comparison to models: }\\
The relative absence of giant cluster members 
makes the photometry sub-optimal for an isochrone fit. 
However, it is certainly useful to compare the observed cluster main sequence 
with a modern theoretical isochrone.
% FS >>>: about the models
Accordingly, we have calculated (and display in Fig.\,3) three suitable 
isochrones, based on the models of Spada et al. (2013) 
and of Yi et al. (2001). 

% Fig.3. The color-magnitude diagram with the isochrone
   \begin{figure}
   \centering
   \includegraphics[angle=-90,trim=0cm 3cm 0cm 0cm,clip=true,width=105mm]{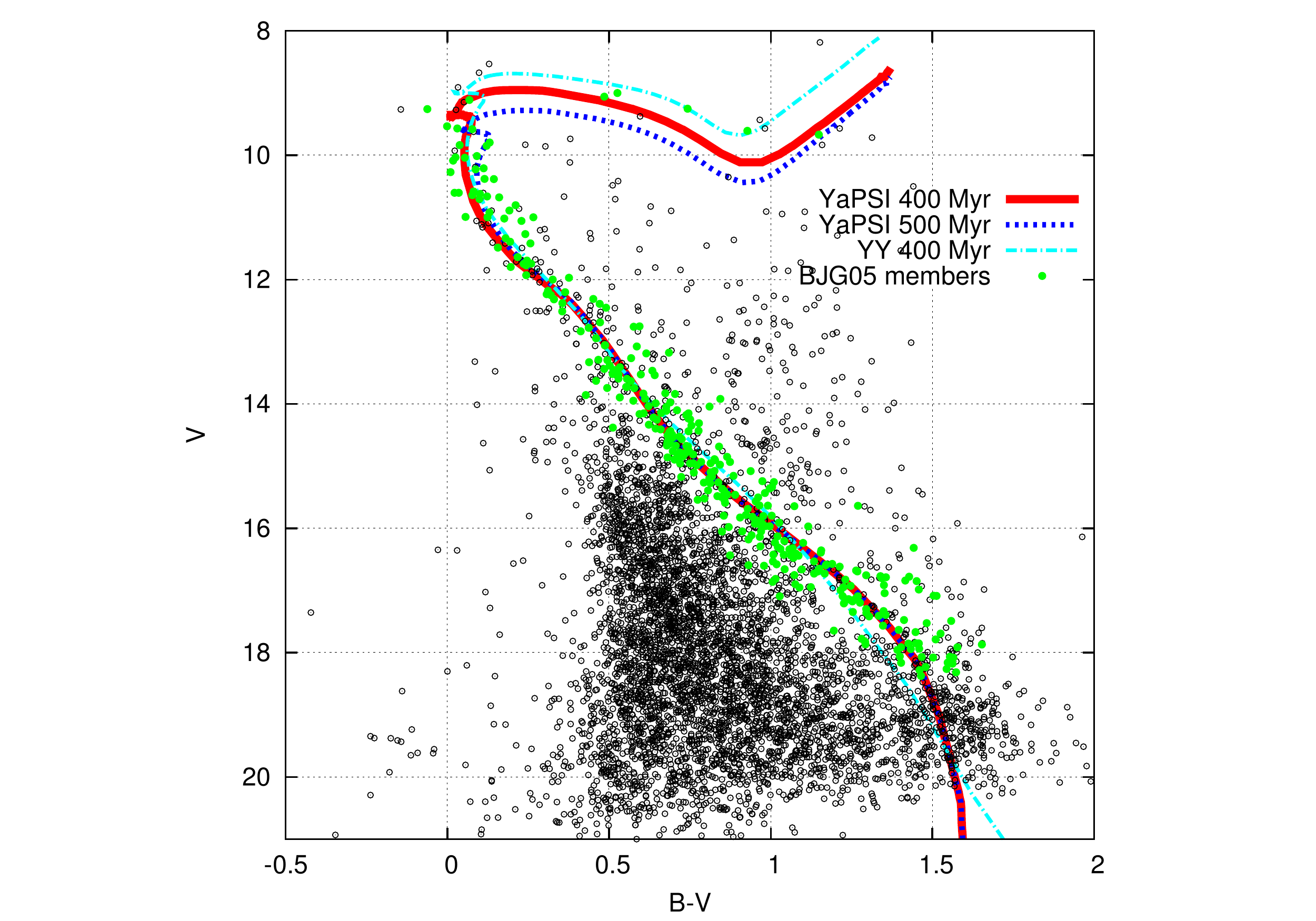}
      \caption{CMD for M\,48 with solar-metallicity isochrones based on Spada et al. (2013; YaPSI) for ages of 400-, and 500\,Myr, and from Yi et al. (2001; YY) for 400\,Myr. The distance modulus and reddening used are $(m-M) = 9\fm3$ and $E(B-V) = 0\fm08$ respectively. 
              }
         \label{fig3}
   \end{figure}

%\newpage
One of the goals of Spada et al. (2013) was to update the Yale-Yonsei ($Y^2$) 
database of stellar models (Yi et al., 2001), with particular attention to the 
input physics relevant for the lower mass regime 
(i.e., $M \lesssim 0.6\, M_\odot$; $(B-V)_0 \gtrsim 1.3$). 
Most notably, the atmospheric boundary conditions are based on the PHOENIX 
model atmospheres (Hauschildt et al., 1999; Allard et al., 2011).
For more details on the micro-physics used in the models, see Spada et al. 
(2013).
% FS <<<

The isochrones have been calculated for solar metallicity, with ages of
400\,Myr (the nominal cluster age), and 500\,Myr. 
These are displayed in Fig.\,3, assuming a distance modulus of 9.3 
($d = 725$\,pc), and a reddening of $E(B-V)=0\fm08$, as determined by BJG05.
We see, unsurprisingly, that the 
400\,Myr and 500\,Myr 
isochrones are essentially identical redward of the A-type stars.
It is also evident from Fig.\,3 that the newer isochrones indeed arguably 
follow the cluster's main sequence somewhat better for redder colors than the 
earlier one (YY; Yi et al. 2001).

\section{Variability}

%{\bf Intro: }\\
The rotational variability and associated periods of the cluster stars provide 
the principal motivation for our study. This goal requires the sustained 
acquisition of high-quality imaging data over a sufficiently long time baseline.
We are particularly fortunate with respect to the baseline because ours is 
$\gtrsim$2\,months long.
This is four times longer than our conservative expectation of $\sim$15\,d for 
the longest periods in our color range for a Hyades-aged cluster, allowing 
multiple phases to be detected for all rotation periods.

%{\bf Selection of exposures: }\\
Because our robotic STELLA telescopes are able to acquire data during both 
good and suboptimal conditions at no additional cost, it is correspondingly 
necessary for us to discard the latter from among the images acquired by the 
telescope.
We have therefore rejected images with bad seeing, those with very elongated 
PSFs (usually caused by tracking errors), those affected by 
lunar proximity, and finally any frames where fewer than 80\% of the stars on 
our reference images were detected. 
Frames with mean deviations greater than 0.2\,mag from these best exposures 
were also discarded.
Most of the last category consist of images acquired under non-photometric 
conditions, as revealed by our weather monitors and standard star observations.
For this particular observing campaign, the final usable high-quality images 
constituted $\sim$40\% of the total acquired.
Fortunately, this selection does not introduce any significant gaps in the time 
series beyond a lunar proximity issue centered on 11\,Mar\,2014, and minor 
weather-related interruptions, mostly in April\,2014.
(The light curve for one of our solar-mass stars, displayed in Fig.\,4,
 demonstrates this continuity in our observations.)
%The month of Mar\,2014 was the better of the two months in many respects.

%{\bf Number of exposures: }\\
%{\bf Light curves: }\\
The aperture photometry from the individual exposures was corrected to a common 
system defined using the best images by cross-identifying $>$1000 well-measured 
stars with $10 < V < 16$ over all four fields and across all the exposures. 
These stars were used as photometric references for making the frame-to-frame 
corrections relative to the best exposures, allowing the construction of 
individual light curves, which were then constructed for all cross-identified 
stars in the field of view of the cluster.
We decided to concentrate our efforts on candidate photometric cluster members. 
Consequently, we then extracted the light curves of $\sim$1300 stars along the 
cluster sequence for careful analysis. 
This is a superset of the cluster members of BJG05 with $B-V > 0.40$, which
is the color range of interest in this work.

% Fig. 4 Example light curve and frequency spectrum
   \begin{figure}
   \centering
   \includegraphics[angle=0,width=95mm]{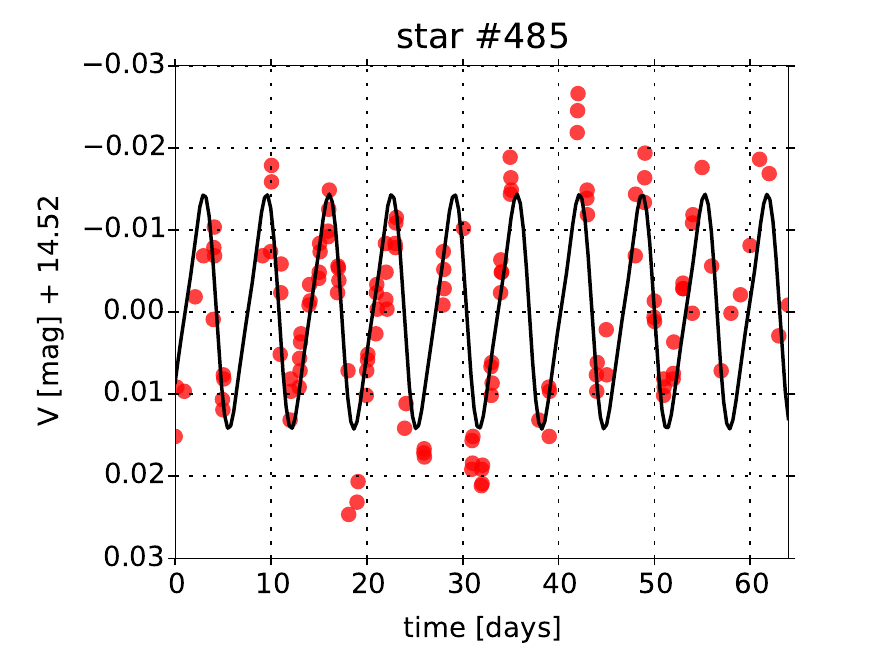}
   \includegraphics[angle=0,width=95mm]{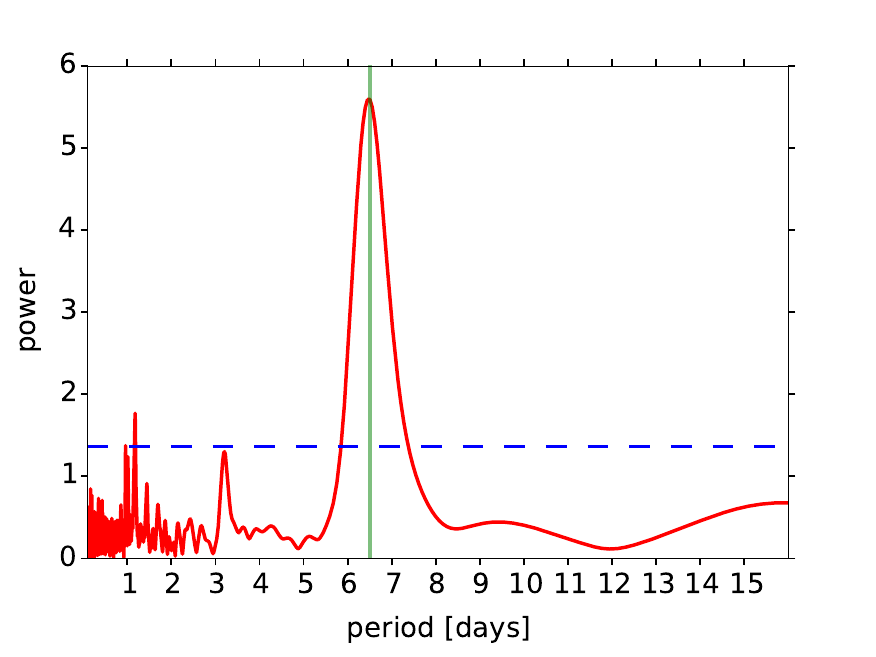}
      \caption{{\it Upper panel:} Light curve for star No.\,485,
a solar-mass star with $(B-V)_0 = 0.664$ in our sample. A periodicity of $6-7$\,d is evident from inspection of the light curve. Successive data points show that the photometry for this particular star is clearly repeatable at a level better than 0.01\,mag. Our other solar-mass rotators have smaller amplitudes of variability. 
{\it Lower panel: } Power spectrum for star No.\,485, showing an unambiguous periodicity of the signal at $P=6.53d$, identified as the rotation period of the star.
%{\bf $\Delta P = 0.08$\,d.}}
              }
         \label{fig4}
   \end{figure}

%{\bf Light curve analysis: }\\
All light curves were subjected to multiple methods of frequency
analysis to identify periodicity -- 
phase dispersion minimization (PDM; Stellingwerf, 1978),
{\sc clean}ed Fourier analysis (Roberts et al. 1987), 
and generalized Lomb-Scargle periodogram (Zechmeister \& Kurster, 2011).
While none of these is completely satisfactory in all cases, we have found that 
for our dataset the {\sc clean} algorithm appears to be the best one overall. 
Lomb-Scargle and PDM are affected more often by the 1\,d or multiple period 
alias. We believe that this behaviour partly comes from our long baseline, 
during which there is sometimes significant spot evolution, the onset/decline 
of spot activity, and in certain cases multiple spot groups. 
{\sc clean} seems to be less sensitive than the other methods to these 
problems, piles up power at the rotational frequency, 
and also allows the window function to be taken into account. 

However, there seems to be no substitute for manual inspection of the 
candidate rotators, taking the raw light curve into account, 
the results from  all three periodicity indicators discussed above and, 
finally, the phased curves.
We found 5$\sigma$ of the noise level in the {\sc clean} spectra to be a 
good threshold for accepting periodicity and have generally adopted this as 
the criterion for a high-quality period (Quality flag = 1). 
However, it cannot be an automatically adopted criterion, so we made the 
ultimate decision manually, and on this basis also listed 
certain lower quality periods (Quality flag = 2), where 
there might be some chance of the listed period being an alias. 
There are therefore some stars with higher peaks that are assigned 
Quality flag = 2, while the rest are considered high quality.
On this basis, we have presented 62 rotation periods, of which 8 are listed
with Quality flag = 2. Experiments show that even our Q = 2 periods
have a Scargle false alarm probability, FAP < 0.01, i.e., a confidence level 
greater than 99\%. 

The light curve for a good example of a solar-mass periodic variable is 
displayed in the upper panel of Fig.\,4.
The locations of successive data points in the light curve show that the
photometry for this particular star is clearly repeatable to a level better
than 0.01\,mag.
A periodicity of $6-7$\,d is obvious in Fig.\,4.
Thanks to the density of our observations, simply plotting the unphased 
light curve usually allows us to verify the approximate periodicity by 
visual inspection, as the upper panel of Fig.\,4 shows.

%{\bf Fourier analysis: }\\
The corresponding Fourier power spectrum for this star, constructed using 
the {\sc clean} algorithm of Roberts et al. (1987) is displayed in the lower
panel of Fig.\,4.
The peak is at $6.53$\,d, which is consequently listed as the rotation period 
of the star.
(Light curves, power spectra, and phased light curves for all the 
periodic candidate cluster rotational variables identified in this study are 
displayed in the online Appendix to this paper in Figs.\,A1-A9.\footnote{The online appendix is available at http://www.edpsciences.org.})
We note that few stars actually need to be phased before their periodicity 
becomes obvious\footnote{Our time baseline is long enough for significant spot evolution. Correspondingly, our phased light curves would look significantly better ``cosmetically'' had we suppressed epochs of low stellar variability. They would also look better had we rephased the light curves to the PDM values, within the error envelope of the {\sc clean} values.}. 
The periods derived range from $1.67$\,d to $13.3$\,d, the latter well within 
the sensitivity of our observational window of $\sim$\,2\,months.
These 62 periods are listed in Table\,2, along with other relevant 
information for each star. 
This includes the rotation period, its error
(the HWHM of the {\sc clean} peak), the variability amplitude (from sinusoidal
fits to the phased light curves), cross-identification with BJG05, the relevant
membership indicator, and relevant notes. The periodic stars considered to be 
cluster members are indicated in the CMD displayed in Fig.\,5.

% Fig.5. The color-magnitude diagram with the rotational variables
   \begin{figure}[ht]
   \centering
   \includegraphics[angle=-90,trim=0cm 3cm 0cm 0cm,clip=true,width=105mm]{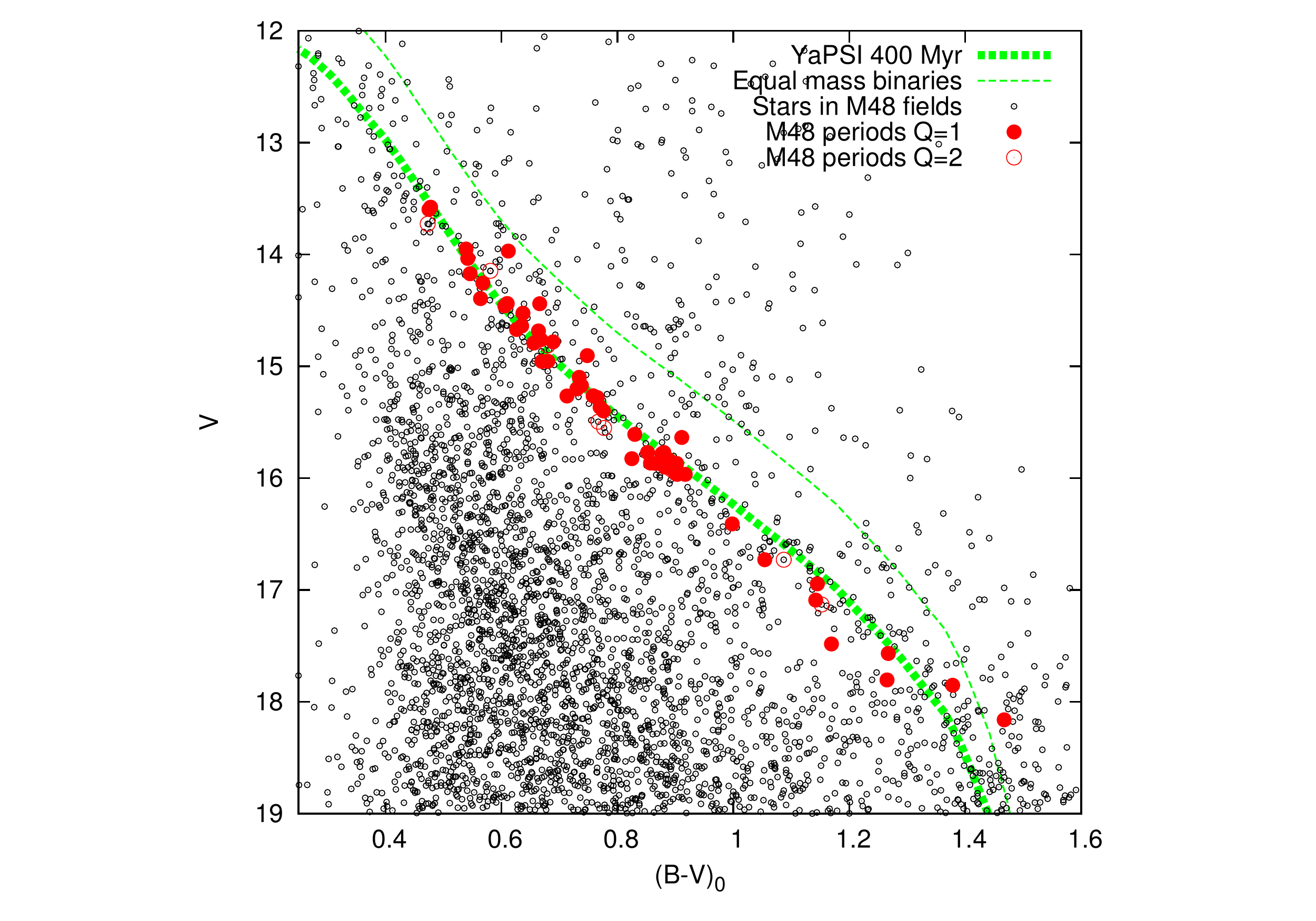}
      \caption{Color-magnitude diagram for M\,48 with the 62 periodic cluster rotational variables highlighted. Solid symbols indicate the 54 high-quality (Q = 1) periods, and unfilled symbols the 8 lower quality (Q = 2) ones. All the periodic cluster member candidates are located on, or relatively near, the cluster sequence. A 400\,Myr YaPSI isochrone and the corresponding equal-mass binary sequence are also displayed.
              }
         \label{fig5}
   \end{figure}

\subsection{Empirical comparisons of the color-period diagram}

%{\bf Intro: }\\
The rotation periods derived by the methods described above have then been 
associated with the photometry discussed earlier in the paper, and the
corresponding CPD has been constructed for the cluster stars, 
as displayed in Fig.\,6.
This diagram is populated by 62 stars with $(B-V)_0$ color range 
0.47\,mag--1.47\,mag, and the period range 1.67\,d--13.1\,d.
The field contamination from the background, as seen in the CMD in Fig.\,2 
suggests that only a couple of these could possibly be non-members.

% Fig.6. The bare color-period diagram
   \begin{figure}
   \centering
   \includegraphics[angle=-90,width=90mm]{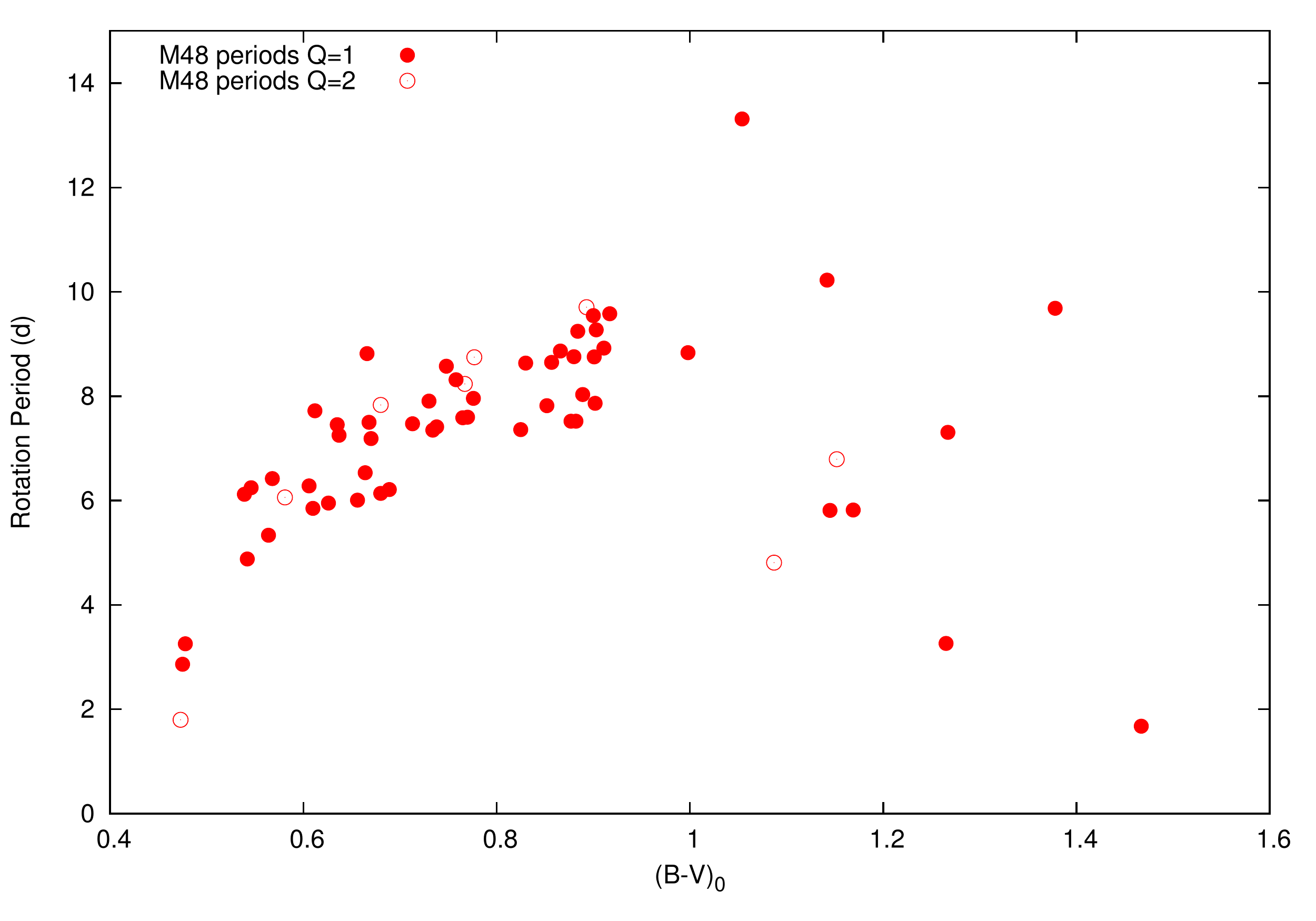}
      \caption{The measured color-period diagram (CPD) for M\,48, showing the 62 periodic stars believed to be cluster rotators, and whose positions in the CMD are displayed in Fig.\,5. We observe a relatively distinct sequence of stars, reminiscent of the Hyades sequence, blueward of $B-V = 1$, and a more scattered distribution of stars redward of this color. The solar-mass stars show no evidence of the C/fast sequence that characterizes ZAMS clusters. 
              }
         \label{fig6}
   \end{figure}

%{\bf Description of CPD: }\\
The bluer (warmer) half of the M\,48 CPD displays a clear 
sequence of stars ranging from short-period $\sim 2$\,d stars at 
$(B-V)_0 \sim 0\fm45$ to $\sim 8$\,d periods at $(B-V)_0 \sim 0\fm9$. 
This is  reminiscent of the situation in the 625\,Myr-old Hyades cluster, as 
discussed below.
There is no evidence of the C/fast sequence of stars with $P \sim$1\,d across
the entire $ 0.5 < (B-V)_0 < 1.5$ color range, characteristically seen in 
younger open clusters such as the Pleiades (Hartman et al. 2010) or M\,35 
(Meibom et al. 2009).
This morphology immediately tells us that M\,48 is older than M\,35
and the Pleiades, without even having to perform a detailed comparison.

% Fig.7. The color-period diagram, compared with the Hyades CPD
   \begin{figure}[h!]
   \centering
   \includegraphics[angle=-90,width=84mm]{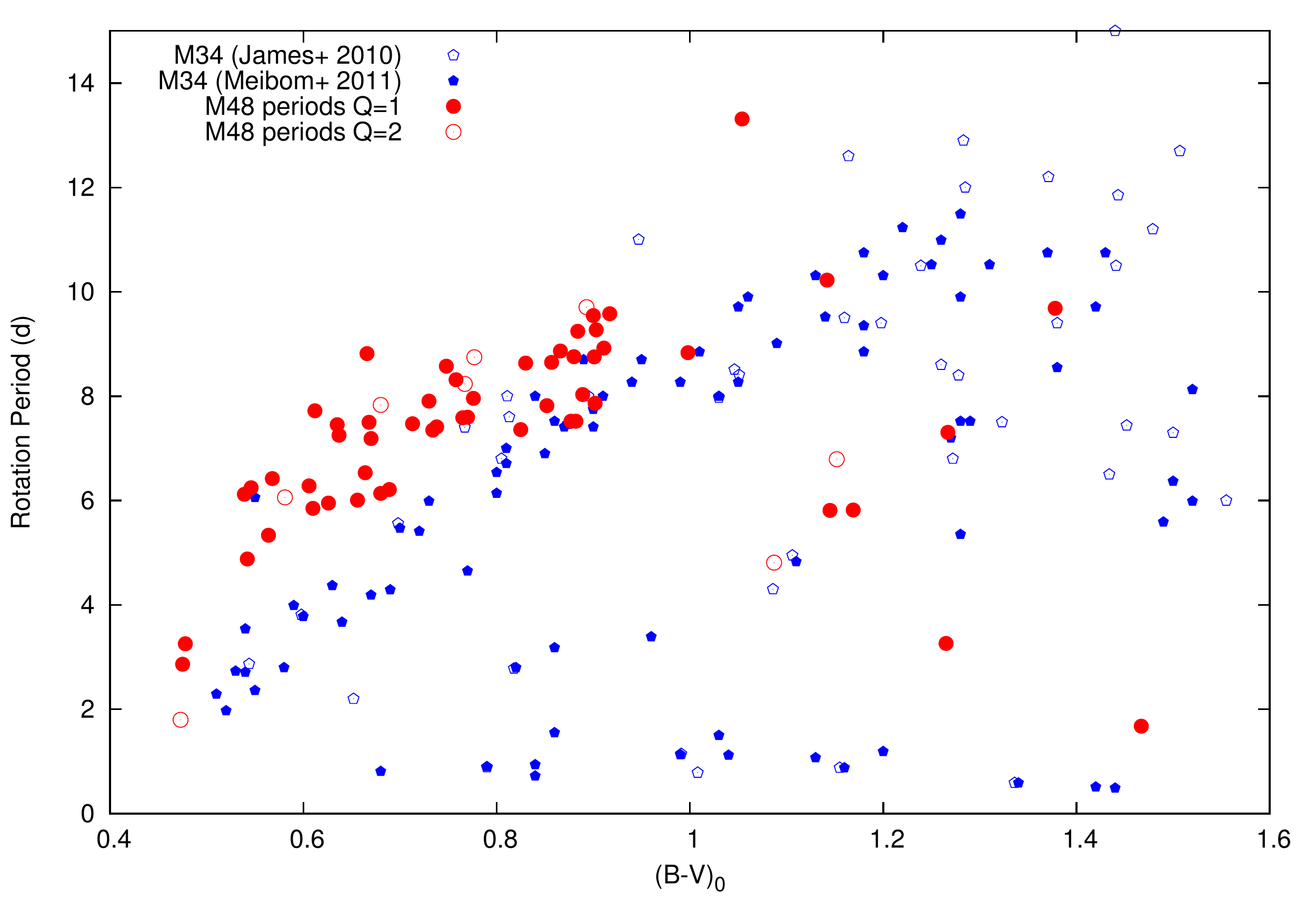}
   \includegraphics[angle=-90,width=84mm]{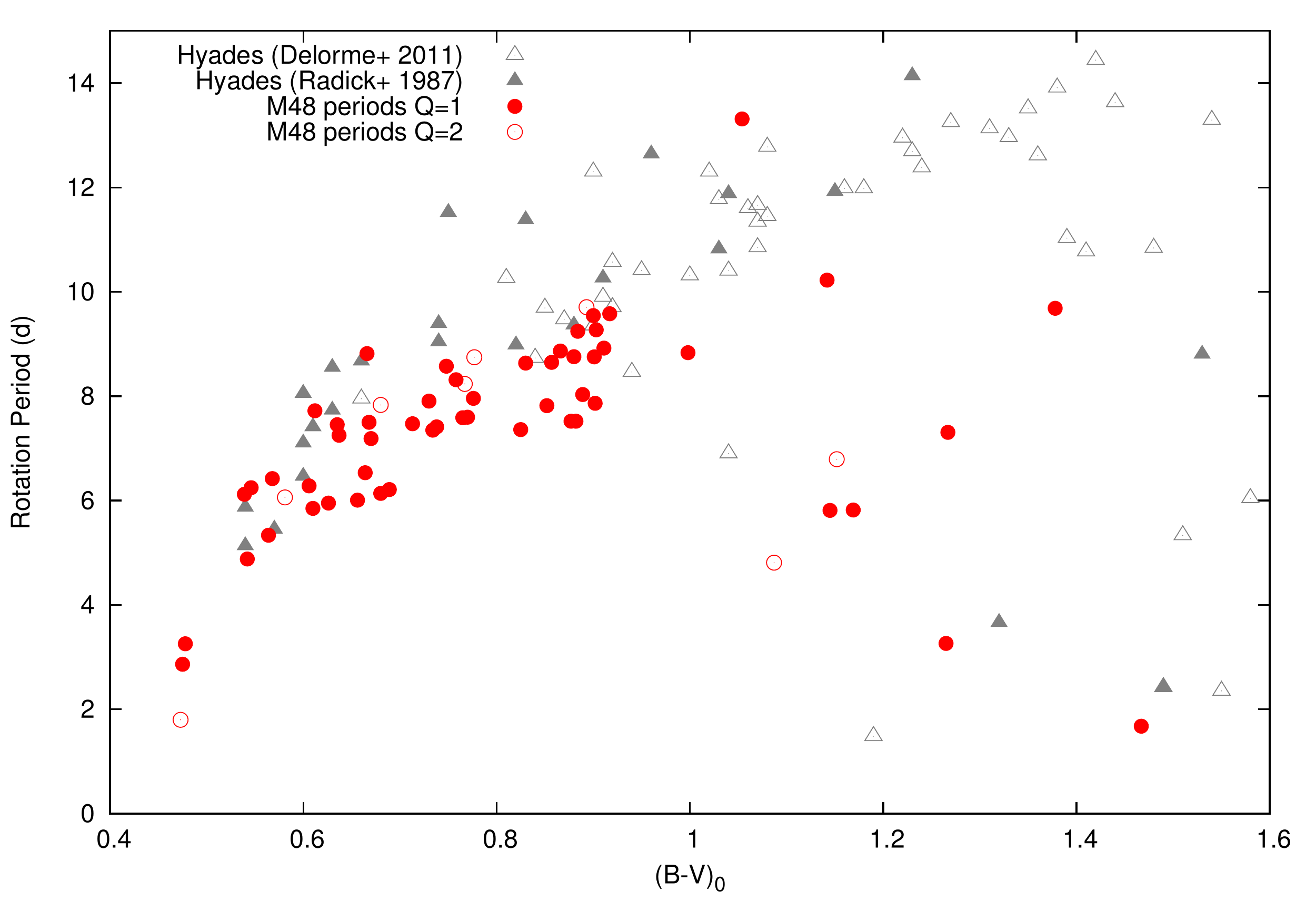}
   \includegraphics[angle=-90,width=84mm]{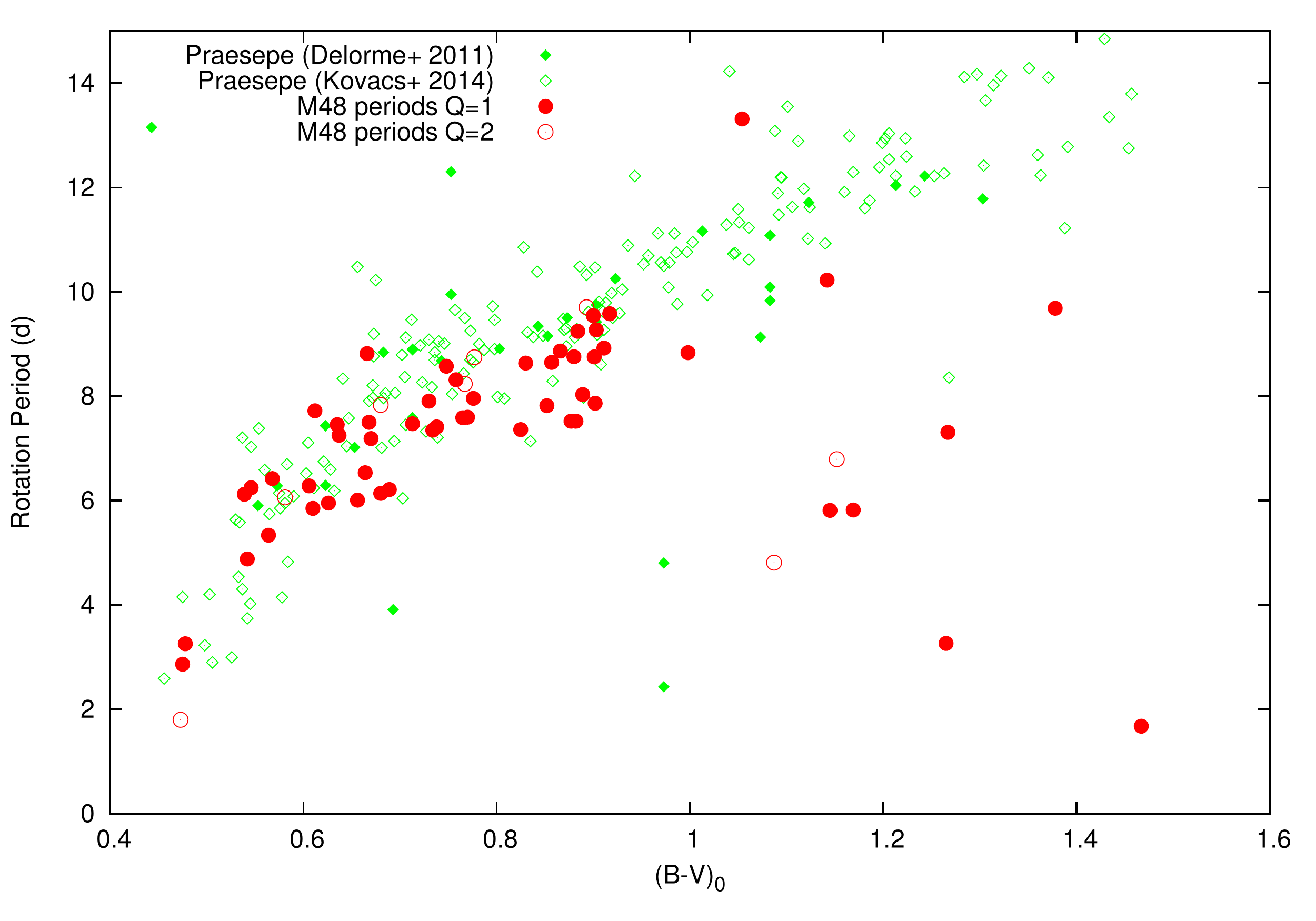}
      \caption{{\it Top panel: } Comparison of the M\,48 CPD (circles, red online) with that of M\,34 (250\,Myr old; pentagons, dark blue online). We see that the M\,48 sequence is clearly above that of M\,34, indicating that M\,48 is older. Another indicator of M\,48's older age is it's lack of fast-rotating stars,which form a sparsely populated sequence in M\,34 (with $P \sim$\,1\,d and $ 0.6 < (B-V)_0 < 1$). 
{\it Middle panel: } Comparison of the M\,48 CPD with that of the Hyades cluster (triangles, gray online). While the rotation periods of similar-mass stars are comparable between the two clusters, indicating that their ages are roughly comparable, the average solar-type M\,48 star (see text) is 0.99\,d below that for the 625\,Myr-old Hyades, indicating that M\,48 is somewhat younger. 
{\it Bottom panel: } Comparison of the M\,48 CPD with that of the 590\,Myr-old Praesepe cluster (diamonds, green online), confirming that M\,48 is again younger. The average solar-type M\,48 star (see text) is closer (0.44\,d), consonant with Praesepe's age being slightly lower than that of the Hyades. 
}
\label{fig7}
   \end{figure}

%{\bf Intro to comparisons, and M34: }\\
A more detailed empirical idea of how the M\,48 cluster fits in with other 
cluster observations can be obtained by comparing the constructed M\,48 CPD 
directly with other open cluster CPDs.
%{\bf Comparison with M34: }\\
Our first empirical comparison is made with the open cluster M\,34, as 
displayed in the top panel of Fig.\,7.
The filled and unfilled symbols for M\,34 represent the rotation periods 
determined by Meibom et al. (2011a) and James et al. (2010). 
Following the original publications, the M\,34 data have been de-reddened by 
$E(B-V) = 0\fm07$ (Canterna et al. 1979).
We see that M\,48 does not possess the fast G- and K-type rotators 
blueward of $(B-V)_0 = 1$ that collectively form an (admittedly ill-defined) 
fast/C sequence in M\,34. 

The M\,48 slow/I sequence is clearly above that of M\,34,
indicating that M\,48 is older than 250\,Myr, the nominal age of M\,34
(Ianna \& Schlemmer, 1993).
The I\,sequences of the two clusters almost overlap at $(B-V)_0=0.9$,
suggesting that it is already time to move beyond models where the dependence
of rotation period, $P$, on age and mass is separable 
[e.g., Barnes\,(2003), Barnes\,(2007), Mamajek \& Hillenbrand (2008), 
Meibom et al. (2011a), Angus et al. (2015)].

%{\bf Comparison with the Hyades CPD: }\\
The corresponding comparison for the Hyades open cluster is displayed in the 
middle panel of Fig.\,7, using the rotation period measurements 
(filled symbols) of Radick et al. (1987), and the more recent measurements 
(unfilled symbols) of Delorme et al. (2011). Taylor (2006) has determined that 
the Hyades reddening is $E(B-V) \le 0\fm01$\,mag, and consequently the $B-V$ 
values from neither of these studies have been dereddened.

We see that the rotation periods of similar-mass stars in M\,48 and in the
Hyades are comparable, indicating that the ages of the two clusters are
roughly comparable. (In their comprehensive study of the Hyades,
Perryman et al. (1998) derived an age of 625\,Myr.)
Closer inspection shows that the average M\,48 rotational sequence in the CPD 
is clearly below the average of the Hyades sequence, telling us that M\,48 is 
somewhat younger. The average difference between the two sequences in the 
well-sampled $0.5 < (B-V)_0 < 0.9$ interval is 0.99\,d.

%{\bf Comparison with other cluster CPDs: }\\
An empirical confirmation of this result is available using the rotation period
determinations for the Praesepe cluster by Delorme et al. (2011) and Kovacs et 
al. (2014). These are plotted in the bottom panel of Fig.\,7 using filled and 
unfilled symbols respectively.
In accordance with the determination of Taylor (2006), the $B-V$ color values
for Praesepe have been de-reddened by $E(B-V) = 0\fm027$.
Praesepe is known to be of very similar age ($590$\,Myr; Fossati et al. 2008) 
to the Hyades. 
Our comparison of the M\,48 and Praesepe CPDs is consonant with this result.
In fact, after comparing the Praesepe and Hyades CPDs, 
Delorme et al. (2011) concluded that Praesepe is slightly younger than the
Hyades, perhaps by about 50\,Myr. 
Our comparison in Fig.\,7 seems to confirm this conclusion in that the 
Praesepe sequence is indeed slightly closer to the M\,48 sequence
than that of the Hyades. The average difference between the two sequences in 
the $0.5 < (B-V)_0 < 0.9$ interval is 0.44\,d, smaller than the Hyades-M48 
difference of 0.99\,d. These empirical comparisons tell us that the
rotational age of M\,48 is almost certainly in the (250, 590)\,Myr interval.

\subsection{Comparison against models with separable (t, M) dependencies}

%{\bf Comparison with B2003, B2007, MH2008, Mei+2011a, Angus+2015: }\\
These new M\,48 rotation periods also permit a useful comparison with prior 
predictions for the locations of stars of its age in the CPD.
Fig.\,8 shows how these data compare with a number of empirical studies that
relied on the rotation period $P$ having separable dependencies on age, $t$,
and stellar mass $M$ (or a suitable mass proxy such as color) --
the original gyrochronology formulation of Barnes (2003), the update in 
Barnes (2007), the modification to that formulation proposed by Mamajek \& 
Hillenbrand (2008), one based largely on the M\,34 cluster data of 
Meibom et al. (2011a), and the recent one of Angus et al. (2015), the last 
based on a very limited cluster dataset, and a small number of field stars 
with ages determined from asteroseismology.
The displayed comparisons all use the prior published age for M\,48 of 400\,Myr.
 Moderate changes to the age used make moderate changes to the displayed curves,
but do not change the overall behavior.

% Fig.8. The color-period diagram, compared with old theory
   \begin{figure}
   \centering
   \includegraphics[angle=-90,width=90mm]{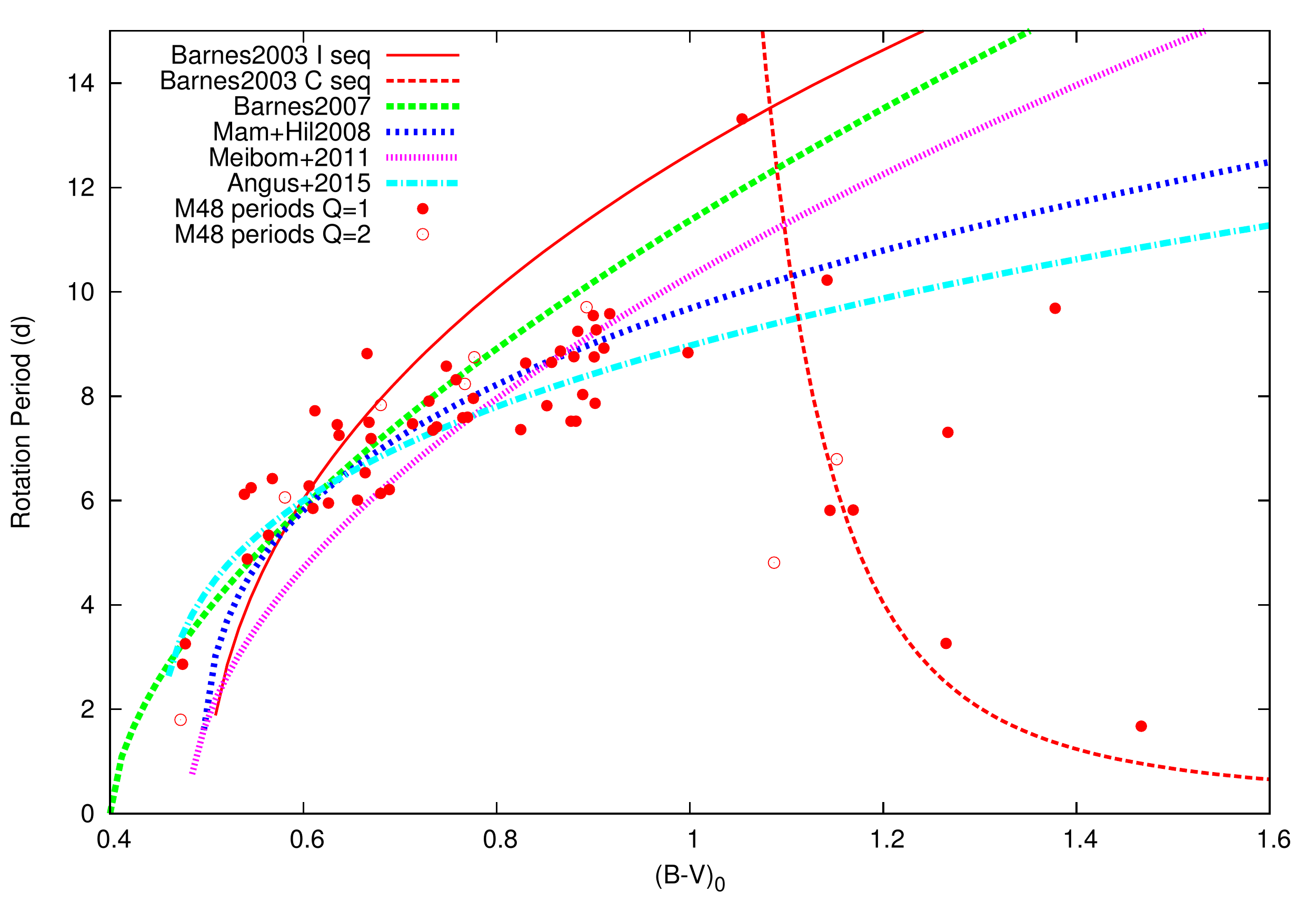}
      \caption{Comparison of the M\,48 CPD with rotational isochrones for models with separable $(t, M)$ dependencies, constructed for an age of 400\,Myr. The M\,48 data points have been de-reddened by $0\fm08$. These models do not satisfactorily capture the detailed morphology of the observations (see text for details). 
%{\bf Gallet+Bouvier?}
              }
         \label{fig8}
   \end{figure}

We observe that all the models displayed here capture, to a certain extent, 
the overall behavior of the M\,48 data blueward of $(B-V)_0 \sim 1$. 
However, there is a potentially serious problem redward of this color value. 
The measured rotation periods seem to decline for redder (lower-mass) stars. 
To a certain extent, this behavior is not unexpected, because these empirical 
formulations of gyrochronology explicitly ignored the fast rotators (with the 
exception of Barnes (2003), which treated them separately as shown in Fig.\,8), 
electing to concentrate attention on the slow/I sequence stars, which are the 
only ones seen in M\,48 blueward of $(B-V)_0 = 1$.

There are also certain differences, obvious in Fig.\,8, with respect to the 
point where these models intersect the color axis, and with the color range 
over which each of these provides a good fit (or not) to the periods. 
(We will show a detailed comparison with a subsequent preferred model below.)
%In this respect, however, we note that the small M\,48 cluster reddening of 
%$E(B-V) = 0\fm08$ does not permit significant room for any left-right 
%adjustment of the measured CPD. 
Finally, it should be noted that the separable empirical models have slightly 
differing dependencies on the age, a fact that is imperceptible in such a 
comparison, being overwhelmed by the mass dependence.

%{\bf Longer-period red stars: }\\
We have not been able to identify any red stars (say $(B-V)_0 \gtrsim 1$) 
that are very slow rotators ($P >$10\,d). While our observations do possess 
the time baseline to identify such stars, any such stars are fainter than 
$V = 16.5$ (see Fig.\,5), where the precision of our photometry declines, 
affecting our sensitivity to such periods if their variability amplitudes are 
small. Our data therefore do not allow us to state conclusively
whether such stars actually exist or not in M\,48.
%(We plan to re-observe these stars carefully in due course.)

\subsection{Comparison with the B2010 model}

%{\bf Ages for the cluster stars: }\\
Our preferred way of interpreting these M\,48 rotation period data is with the 
gyrochronology formulation of Barnes (2010), hereafter B10, based on the 
rotational evolution ideas described in Barnes \& Kim (2010). 
A significant part of this preference arises from the simplicity of this model,
where only two dimensionless constants, $k_C$ and $k_I$, are required to 
specify the two relevant spindown timescales for all fast and slow cool stars,
and hence of the entire main sequence rotational evolution of cool stars. 
The B10 formulation describes the behaviors of both the slow/I-type stars and 
the fast/C-type stars in a single model, treating them symmetrically\footnote{Brown (2014) has called this the `Symmetric Empirical Model.'}.
Furthermore, the rotational isochrones constructed using this model are 
better than those from other models in matching the measured color-period 
diagram of the 2.5\,Gyr-old cluster NGC\,6819, studied by Meibom et al. (2015).
And we show below that this model describes the morphology of the M\,48 
CPD in considerable detail.

This B10 model provides the age of an individual star as an explicit function 
of its rotation period (and a mass variable), albeit with more complexity than 
the separable empirical models discussed above. One can then take a suitable 
average over the cluster stars to derive the corresponding cluster age.
The procedure first requires the calculation of the (global) convective 
turnover timescale, $\tau$, for each star from its de-reddened $B-V$ color. 
This is simply accomplished by interpolation from Table\,1 in Barnes \& Kim
(2010). Then one applies equation (32) from B10,
\begin{equation}
t = \frac{\tau}{k_C} {\rm ln}\frac{P}{P_0} + \frac{k_I}{2 \tau} \left( P^2 - P_0^2\right).
\end{equation}
This is an explicit expression for the age $t$ of the star, in terms of its
measured rotation period $P$, and $\tau$, the latter a proxy for its mass or
color. Here, $k_C = 0.646$\,Myr/d and $k_I = 452$\,d/Myr are two dimensionless 
constants whose values are decided by the totality of the open cluster rotation
period data, and the solar datum; we have simply adopted unchanged the values
proposed in B10, particularly in view of the model's ability to reproduce well 
the CPD of the 2.5\,Gyr-old cluster NGC\,6819, as discussed in Meibom et al. 
(2015).
$P_0$ is the initial rotation period, set to $P_0 = 1.1$\,d, following
B10, but we will also allow it to vary within certain bounds below,
as allowed by observations, and as described in B10.
(It is also possible to use other turnover timescales, but then one must be
 careful to recalibrate the constants $k_C$ and $k_I$ to match the totality
 of the open cluster and solar rotational data, as discussed in B10.) 

% Fig.9. The histogram of age
   \begin{figure}
   \centering
   \includegraphics[angle=-90,width=90mm]{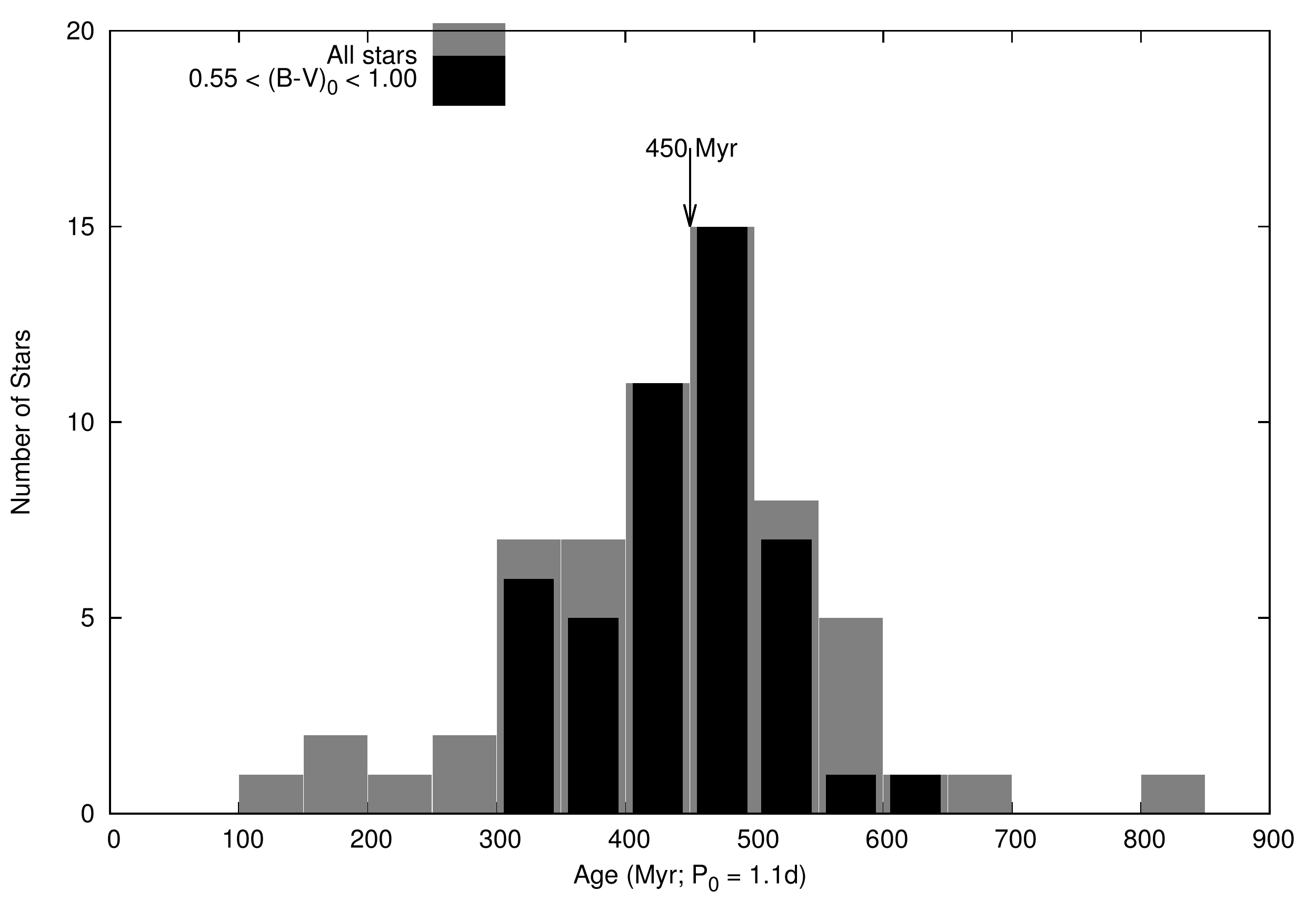}
      \caption{Histograms of gyro-age for the individual M48 stars, calculated using the B10 rotational model and assuming an initial period of 1.1\,d. The gray histogram shows all 62 stars, while the black histogram is for a sample restricted to the 46 stars in the particularly well-measured $0.55 < (B-V)_0 < 1.00$ region. The larger sample gives a median age of 451\,Myr (mean = 441\,Myr, S.D. = 117\,Myr = 26\%), and the smaller one gives similar values, with a median age of 457\,Myr (mean = 447\,Myr, S.D. = 70\,Myr = 16\%.) This large age dispersion mostly arises from having to assume a single initial period, $P_0$ (= 1.1\,d) for all stars. 
              }
         \label{fig9}
   \end{figure}

Provided that the measured period is greater than $P_0$, each measured
rotation period $P_i$ results in a corresponding gyro-age $t_i$, which is 
plotted in the histograms in Fig.\,9 for $P_0 = 1.1$\,d.
The gray histogram shows the distribution for all 62 stars with
measured rotation periods. 
We obtain a distribution that is sharply peaked at a 
median value of 451\,Myr, and mean value of 441\,Myr 
but with relatively wide wings, giving a standard deviation of 117\,Myr 
(= 26\%).

A significant portion of this relatively large dispersion can be traced to a
number of outliers. For instance, the outlier at $t_i = 840$\,Myr is No.1456, 
the one above all the other stars in the CPD, with $(P, (B-V)_0) = (13.3$\,d, 
$1.054$). Because it lies on the cluster sequence in the CMD, there is no
good reason to discard it at the present time.
The other outliers are mostly either very blue or very red stars (defined for 
this study as lying outside the  $0.55 < (B-V)_0 < 1.00$ color range.)

The black histogram in Fig.\,9 shows the distribution for the 46 
stars in this restricted $0.55 < (B-V)_0 < 1.00$ color range, where the 
I\,sequence is particularly well-defined. 
This color region contains the Sun and many well-studied open clusters, and is 
known from the rotation period observations for NGC\,6819 ($2.5$\,Gyr) by 
Meibom et al. (2015) to be well-calibrated. 
In the corresponding histogram, many of the outliers disappear to reveal a 
distribution where $38/46 = 83$\% of the stars are confined to the 
$t = [350,550]$\,Myr region.
An M\,48 cluster gyro-age outside this range is essentially ruled out.
The formal mean age for this restricted distribution is 
$\overline{t} = 447$\,Myr (median = 457\,Myr),
but with a smaller standard deviation $\sigma_t =$ 70\,Myr (= 16\%).
These results enable us to confirm the $400\,\pm100$Myr isochrone age of this 
cluster, and to propose a mean rotational (gyrochronology) age of $450$\,Myr 
for M\,48.

% Fig.10. The color-period diagram, compared with theory
   \begin{figure}
   \centering
   \includegraphics[angle=-90,width=90mm]{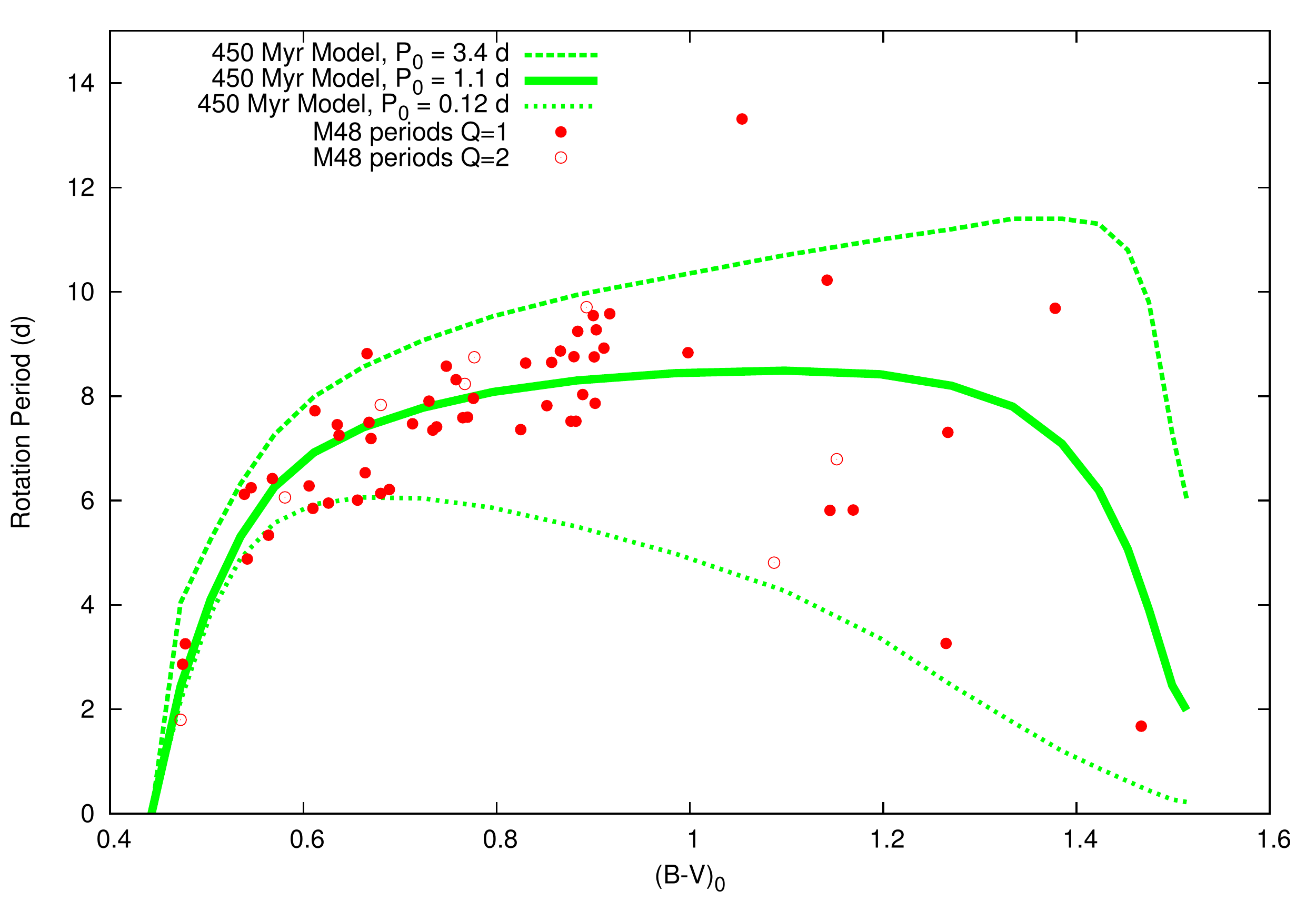}
      \caption{Theoretical rotational isochrones for $450$\,Myr, constructed following Barnes (2010), are compared with the M\,48 CPD. The three curves from top to bottom respectively correspond to initial periods of $3.4$\,d, $1.1$\,d, and $0.12$\,d, representing the range allowed by ZAMS cluster observations. Almost all the data points are consistent with a single rotational age of $450$\,Myr.
              }
         \label{fig10}
   \end{figure}

We now show that much of the above scatter arises from intrinsic astrophysical 
variations in the initial rotation periods, so that it is likely that the 
uncertainty on the gyro-age of M\,48 is actually significantly smaller than 
the 70\,Myr standard deviation for the best-measured stars. 
For open clusters that are significantly younger than the Hyades, it is 
well-known, and expected, that the rotation periods of cool stars will not 
necessarily have converged to a single sequence. 
Indeed, in ZAMS clusters, two distinct sequences are sometimes observed,
as discussed earlier, and as is visible in the CPD for M\,34
(see Fig.\,7). 
In the M\,48 CPD, stars with masses greater than $0.9 M_{\odot}$ ($B-V_0 < 1.0$) 
have essentially converged to a single sequence, while lower mass stars most 
certainly have not.

There is a simple way to understand the dispersion in the age distribution
derived above.
Given an age $t$, and using a given initial period $P_0$, one can solve 
equation (1) numerically\footnote{For a given $P, P_0$, and age $t$, one can also obtain a solution for $\tau$ analytically, as shown in Barnes (2010), by simply solving a quadratic equation.} 
to associate a value $P$ to every $\tau$ value of interest, or equivalently the 
value of $(B-V)_0$. 
Such a curve is the {\it isochrone} for that initial period, since it is the 
locus of all such equal-age points.
The 450\,Myr isochrone for $P_0 = 1.1$\,d is displayed in Fig.\,10 with 
the thick central  line (solid green).
Barnes (2010) found that the lower- and upper envelopes of the initial period 
distribution could be reasonably set at 0.12\,d and 3.4\,d respectively.
Carrying out the corresponding calculations for $P_0 = 0.12$\,d and 
$P_0 = 3.4$\,d  with the same age of 450\,Myr yields the lower (dotted 
green) and upper lines (dashed green) respectively.
(Other intermediate initial period values in this range would provide curves
 that are bounded by these.)

This collection of curves can be viewed as representing the total rotational 
isochrone for M\,48, providing a narrow sequence at the blue/solar-like end, 
and a wider sequence for lower mass stars.
We observe that this range of initial conditions explains almost all the
scatter in the rotation period measurements as a function of mass, because
only one of our 62 rotation periods lies significantly outside 
the range of these isochrones. 
Consequently, with $P_0$ allowed to vary within the astrophysical limits 
permitted by ZAMS observations, 
{\it almost all our stars are consistent with a single age of 450\,Myr.} 
This fact is consistent with the current belief that open clusters are 
simple stellar populations, and that they are describable as single-age 
populations.

% Discussion of cluster age error

%These facts inform us that the age distributions we have constructed (and 
%displayed in Fig.\,9) actually relate to a single underlying cluster age 
%(estimated by its mean), and that it is permissible to quote the standard 
%error (S.E.) of the mean, 
%$\sigma_t/\sqrt{\rm Number\,of\,stars} = 120/\sqrt{74} \approx 15$\,Myr 
%as its error.
%We therefore propose a gyro-age for the open cluster M\,48 of 440$\pm$15\,Myr 
%(S.E. of the mean) $\pm$ (systematic errors in gyrochronology), 
%in agreement with the isochrone age of 400$\pm$100\,Myr.

We now ask how well we know the gyro-age of the cluster as a whole.
The histogram of ages (displayed in Fig.\,9) shows that moving $\pm 50$\,Myr 
off the mean age halves the bin occupancies.
Bins beyond these ages are occupied by stars for which 1.1\,d is not a good
estimate for the initial period, as Fig.\,10 shows.  
We therefore construct 400\,Myr and 500\,Myr rotational isochrones, and
display them in Fig.\,11 (dotted blue and dashed red, respectively),
in an analogy with classical isochrones. 
Neither of these can be considered a reasonable fit to the M\,48 CPD. 
In particular, the 400\,Myr and 500\,Myr isochrones for $P_0 = 1.1$\,d 
pass below and above the core of the rotation period distribution with
$0.6 < (B-V)_0 < 0.9$.
The $P_0 = 3.4$\,d isochrones are not particularly informative.
However, the $P_0 = 0.12$\,d isochrones for $400-$ and $500$\,Myr also pass 
respectively below and above the group of the fastest cluster rotators at 
near-Solar color (and maybe even redder values), making them significantly
worse fits to the fast edge of the rotation period distribution.
We therefore consider 50\,Myr to be a reasonable estimate of the uncertainty
on the gyro-age of M\,48, which we suggest is $450\pm50$\,Myr.

% Fig.11. Error in the rotational age
   \begin{figure}
   \centering
   \includegraphics[angle=-90,width=90mm]{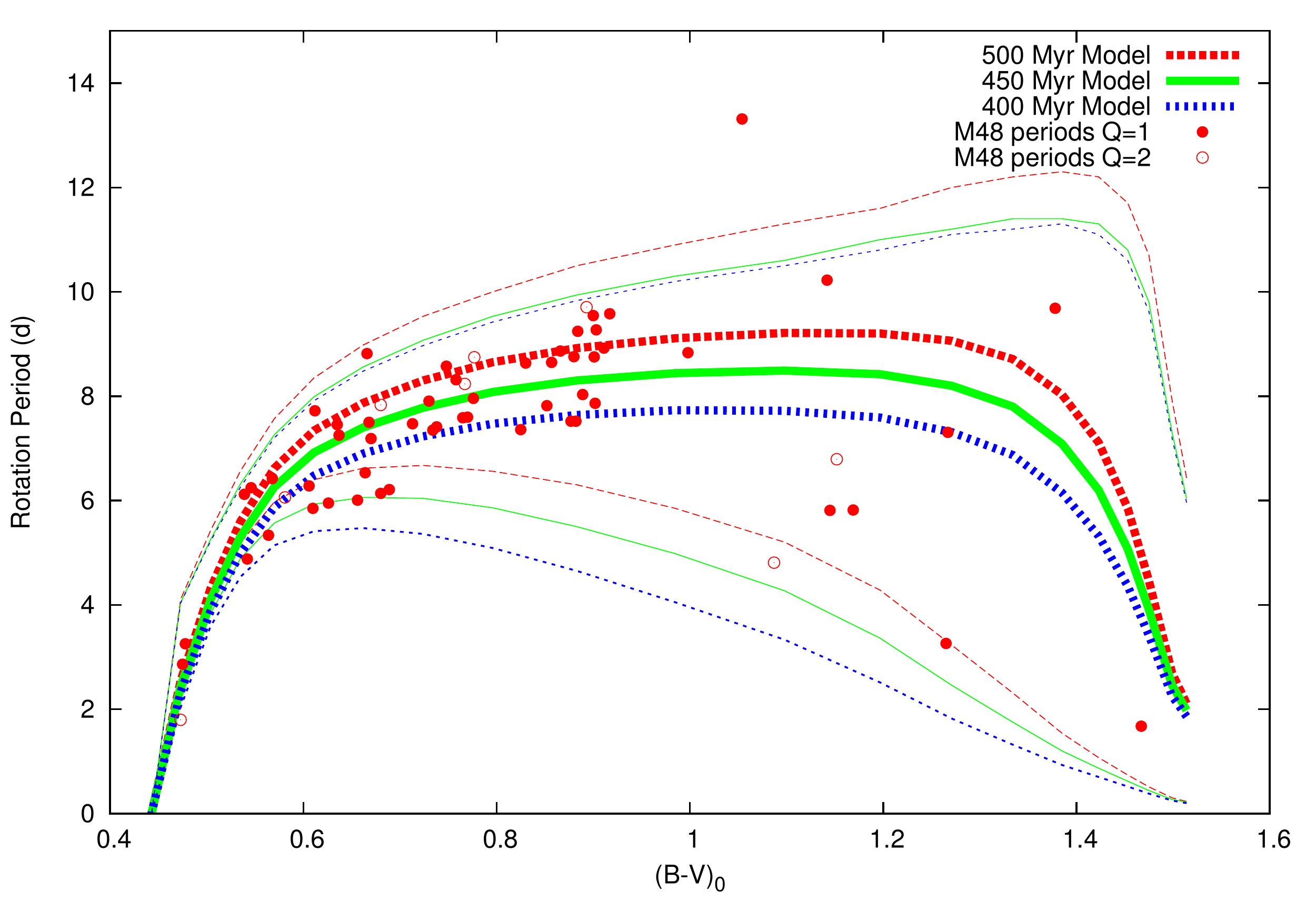}
      \caption{CPD for M\,48 compared with younger ($400$\,Myr; dotted, blue online) and older ($500$\,Myr; dashed, red online) rotational isochrones. Neither the $400$\,Myr nor the $500$\,Myr isochrone provide a good match to the data, passing significantly below or above the core group of stars with $0.6 < (B-V)_0 < 0.9$. The cluster's rotational age is clearly within this (400, 500)\,Myr interval. (For clarity, the (central) $P_0 = 1.1$\,d isochrones for the three separate ages are emphasized.)
              }
%         \label{CPD_err}
   \end{figure}

Another approach is the following.
Ignoring their individual membership of M\,48 and treating all the 62 measured 
stars independently gives the histogram displayed in Fig.\,9, with its standard
deviation of 117\,Myr. 
Clearly, the uncertainty on the mean age of the cluster
must be much lower than the uncertainty of an individual star.
The standard error (S.E.) of the mean is formally
$\sigma_t/\sqrt{\rm Number\,of\,stars} = 117/\sqrt{62} \approx 15$\,Myr. 
But taking this to represent the error in cluster age seems unjustified, in 
view of the possibility of residual systematic age errors arising from our
incomplete understanding of rotational stellar evolution.

To estimate possible systematic errors, we constructed equivalent age 
distributions for the comparison clusters displayed in Fig.\,7, again using 
the $ 0.55 < (B-V)_0 < 1.0 $ interval. 
These give mean ages of 260, 522, and 584\,Myr for M\,34, Praesepe, and the 
Hyades, respectively, which suggest that if the corresponding isochrone ages 
of 250, 590, and 625\,Myr, are taken as absolute truth, then systematic errors 
in the B10 version of gyrochronology could contribute at the level of 
$\sim 40$\,Myr to the M\,48 age uncertainty. Then adding, in quadrature, the 
$15$\,Myr internal error of the mean for M\,48 gives $44$\,Myr.
This suggests that the $\pm 50$\,Myr value quoted above is not unreasonable.

In closing, it is worth noting that an analysis like the one above might not 
be as easily accomplished for a much younger cluster, where the lower-mass 
stars are still on the pre-main sequence (and the demands on rotational 
evolution models are more severe). Indeed Cargile et al. (2014) appear to have 
experienced some difficulties in adapting the Barnes (2010) formulation for 
the $\sim 150$\,Myr-old Blanco\,1 open cluster, and resorted to the separable 
$(t, m)$ formulations for $P = P (t,m)$ to interpret that cluster's 
CPD.

\section{Conclusions}

We have performed a two-month-long photometric time series campaign on the
southern open cluster M\,48 (NGC\,2548) using the AIP's STELLA\,I robotic 
imaging telescope and associated WiFSIP 4K imager, located in Tenerife.
We also acquired photometry in the Johnson B and V bands to a depth of 
$V \sim 20$ for the $44^{\prime} \times 44^{\prime}$ region centered on the 
cluster.
A relatively clear cluster sequence is visible in our photometry and largely
coincides with the astrometric and photometric candidate members identified by
prior work on the cluster.
This sequence in the CMD is followed closely by a theoretical isochrone
and is also closely matched by the rotational variables.

We constructed light curves populated with $\sim$120 data points each 
and with no serious data gaps over the two-month observing baseline. Our 
time-series photometry is repeatable at the 3\,mmag level for F-G stars, 
with the uncertainty increasing as expected for fainter stars.
We successfully derived rotation periods for 62 cool photometric cluster member 
stars, 54 of which are classified as higher-quality, and 8 are lower-quality. 
These rotation periods and the associated colors of the stars define the M\,48 
cluster's CPD, the rotational equivalent of the CMD.
In the CPD, these periods delineate a clear sequence blueward 
of $(B-V)_0 = 1.0$; redward of this point, the rotation period distribution has 
a significantly larger scatter that is likely astrophysical.

While comparable with the rotational sequences in the 625\,Myr-old 
Hyades open cluster and the 590\,Myr-old Praesepe cluster, this sequence 
lies below both, empirically demonstrating that M\,48 is younger.
Likewise, direct comparison with the CPD for the 250\,Myr-old
M\,34 open cluster shows that M\,48 is older.

We constructed the distribution of gyro-ages for the cluster stars, 
finding one that is sharply peaked at $450$\,Myr, but with relatively wide 
wings, giving a standard deviation of $117$\,Myr (= 26\%), and suggesting 
that this
precision for comparable field star rotators is reasonable and attainable.

We showed that the known (astrophysical) contribution from initial period 
variations on the ZAMS accounts for much of the scatter in the rotational
ages, so that almost all the measured rotation periods are actually 
consistent with a single cluster age of $450$\,Myr. 
For the cluster as a whole, the age uncertainty is about 50\,Myr.
We therefore propose a mean cluster age from gyrochronology for M\,48 of
$450 \pm 50$\,Myr.

In sum, this study has added another open cluster of 
intermediate age to the canonical literature.

%__________________________________________________________________

\begin{acknowledgements}
      This work was supported by the German
      \emph{Deut\-sche For\-schungs\-ge\-mein\-schaft, DFG\/} project
      number DFG STR645/7-1.
      This work is based on data obtained with the STELLA robotic telescopes
      in Tenerife, an AIP facility jointly operated by AIP and IAC.
      Detailed comments from an anonymous referee are also appreciated.
\end{acknowledgements}

%-------------------------------------------------------------------

\newpage
\begin{footnotesize}
\onecolumn
% Table2
\begin{longtab}
\begin{longtable}{rccccccccll}
\caption{\label{tab:rotators}Rotation periods of the M\,48 stars.}\\ 
\hline\hline
Id & V   & B--V & P    & Perr & amp & mem & Q & BJG \# & Notes\\
   & mag & mag  & days & days & mag &     &   &        &      \\
\hline
\endfirsthead

\caption{continued.}\\
\hline\hline
Id & V   & B--V & P    & Perr  & amp &  mem & Q  & BJG \# & Notes\\
   & mag & mag  & days & days  & mag &      &     &       &      \\
\hline
\endhead
\hline
\endfoot
%%%%
 284 & 13.577 & 0.56 & 3.25 & 0.08 & 0.013 & -- & 1 &  & Variable in 2nd half\\ % Bin seq?; 2nd half good
 287 & 13.597 & 0.56 & 2.86 & 0.10 & 0.018 &  M & 1 & BJG 3157 & \\ % Pwrat 2.86d also; (2 spt gps)
 303 & 13.726 & 0.55 & 1.80 & 0.02 & 0.007 &  M & 2 & BJG 841 & Low Amplitude\\ % Pwrat 4.47 also; PDM picks harm.
 332 & 13.951 & 0.62 & 6.12 & 0.41 & 0.037 & -- & 1 &  & \\ % 
 336 & 13.970 & 0.69 & 7.72 & 0.56 & 0.030 & -- & 1 &  & \\ % Binary MS?
 343 & 14.034 & 0.62 & 4.88 & 0.22 & 0.015 &  M & 1 & BJG 3584 & \\ % Amp<0.01
 363 & 14.146 & 0.66 & 6.06 & 0.42 & 0.017 &  M & 2 & BJG 2049 & 2 spot groups, PDM period\\ % :pwrat3.02max Caution! phased period changed from 5.83 to 6.06d
 368 & 14.172 & 0.63 & 6.24 & 0.35 & 0.022 & -- & 1 &  & \\ % 
 386 & 14.255 & 0.65 & 6.42 & 0.35 & 0.014 &  M & 1 & BJG 2109 & \\ % 
 407 & 14.395 & 0.64 & 5.33 & 0.26 & 0.020 &  M & 1 & BJG 3680 & \\ % Poor ltcrv, period clear
 418 & 14.439 & 0.69 & 5.85 & 0.36 & 0.019 & -- & 1 &  & \\ % :Maxpwrat 0.85d see freq
 421 & 14.440 & 0.75 & 8.82 & 0.64 & 0.012 &  M & 1 & BJG 2037 & \\ % 
 425 & 14.459 & 0.69 & 6.28 & 0.44 & 0.009 &  M & 1 & BJG 1614 & \\ % 
 437 & 14.525 & 0.72 & 7.25 & 0.72 & 0.019 &  M & 1 & BJG 3583 & \\ % 
 467 & 14.640 & 0.71 & 7.45 & 0.57 & 0.014 & -- & 1 &  & \\ % Pwrat 3.4d also
 482 & 14.671 & 0.71 & 5.95 & 0.32 & 0.029 &  M & 1 & BJG 2975 & \\ % 
 485 & 14.683 & 0.74 & 6.53 & 0.42 & 0.028 &  M & 1 & BJG 4781 & \\ % 
 501 & 14.762 & 0.75 & 7.50 & 0.52 & 0.018 &  M & 1 & BJG 521 & \\ % 
 507 & 14.781 & 0.77 & 6.21 & 0.36 & 0.018 & -- & 1 &  & \\ % OK
 511 & 14.792 & 0.74 & 6.00 & 0.23 & 0.020 &  M & 1 & BJG 3846 & \\ % Pwrat 0.86d max     
 517 & 14.812 & 0.76 & 7.83 & 1.73 & 0.011 &  M & 2 & BJG 3927 & P$\sim 7$\,d possible, DR?\\ % DR?, shorter P~7d also possible
 540 & 14.904 & 0.83 & 8.57 & 0.72 & 0.012 &  M & 1 & BJG 2600 & \\ % Pwrat 4.11d also
 555 & 14.955 & 0.76 & 6.14 & 0.37 & 0.023 &  M & 1 & BJG 1608 & \\ % 
 556 & 14.956 & 0.75 & 7.19 & 0.38 & 0.018 &  M & 1 & BJG 4335 & \\ % Maxpwrat f+1
 602 & 15.098 & 0.81 & 7.35 & 0.44 & 0.032 &  M & 1 & BJG 3935 & \\ % 
 620 & 15.172 & 0.82 & 7.41 & 0.57 & 0.026 &  M & 1 & BJG 3168 & \\ % 
 633 & 15.201 & 0.81 & 7.90 & 0.80 & 0.019 &  M & 1 & BJG 1641 & \\ % Pwrat 3.75d also   
 649 & 15.264 & 0.84 & 8.31 & 0.57 & 0.012 &  M & 1 & BJG 1917 & \\ % ltcrv poor, per clear
 652 & 15.266 & 0.79 & 7.47 & 0.58 & 0.016 &  M & 1 & BJG 3964 & \\ % Amp=0.01 Below MS?
 657 & 15.279 & 0.85 & 7.59 & 0.48 & 0.036 & -- & 1 &  & \\ % Pwrat 3.82d also
 699 & 15.365 & 0.85 & 7.60 & 0.79 & 0.029 &  M & 1 & BJG 2997 & \\ % Poor ltcrv, period clear
 713 & 15.399 & 0.86 & 7.96 & 0.64 & 0.032 &  M & 1 & BJG 2704 & \\ % 
 752 & 15.496 & 0.85 & 8.23 & 0.51 & 0.014 &  M & 2 & BJG 3852 & PDM breaks ambiguity\\ % 
 772 & 15.552 & 0.86 & 8.75 & 0.79 & 0.011 &  M & 2 & BJG 3597 & LS ambiguous\\ % Pwrat 1.16d max
 796 & 15.609 & 0.91 & 8.63 & 0.61 & 0.022 & -- & 1 &  & \\ % 
 807 & 15.637 & 0.99 & 8.92 & 0.63 & 0.018 &  M & 1 & BJG 915 & 2 spot groups\\ % 
 862 & 15.771 & 0.96 & 8.75 & 0.72 & 0.040 & -- & 1 &  & \\ % 
 864 & 15.771 & 0.93 & 7.82 & 0.54 & 0.065 & -- & 1 &  & \\ % 
 872 & 15.782 & 0.96 & 7.52 & 0.45 & 0.049 & -- & 1 &  & \\ % 
 898 & 15.826 & 0.90 & 7.36 & 0.54 & 0.036 &  M & 1 & BJG 1828 & \\ % 
 909 & 15.838 & 0.97 & 8.03 & 0.55 & 0.019 & -- & 1 &  & \\ % 
 920 & 15.866 & 0.94 & 8.65 & 0.67 & 0.017 & -- & 1 &  & \\ % 
 921 & 15.868 & 0.98 & 7.86 & 0.47 & 0.022 & -- & 1 &  & \\ % P in 2nd half only
 923 & 15.869 & 0.95 & 8.87 & 0.69 & 0.022 & -- & 1 &  & \\ % 
 931 & 15.889 & 0.98 & 8.75 & 0.68 & 0.025 & -- & 1 &  & \\ % 
 935 & 15.897 & 0.96 & 9.24 & 0.75 & 0.037 & -- & 1 &  & \\ % Pwrat 4.53d also  cmdmem
 937 & 15.902 & 0.96 & 7.52 & 0.50 & 0.035 &  M & 1 & BJG 2546 & \\ % 
 954 & 15.922 & 0.97 & 9.70 & 0.79 & 0.013 &  M & 2 & BJG 1703 & Low Amplitude\\ % Pwrat f+1 also
 969 & 15.956 & 0.98 & 9.55 & 0.90 & 0.021 &  N & 1 & BJG 2435 & \\ % Pwrat 4.53also 2sptgps
 974 & 15.966 & 0.98 & 9.27 & 0.68 & 0.016 &  M & 1 & BJG 2033 & \\ % 
 975 & 15.967 & 1.00 & 9.58 & 0.62 & 0.028 &  M & 1 & BJG 578 & \\ % Max pwrat 1.12d
1227 & 16.409 & 1.08 & 8.83 & 0.72 & 0.064 &  M & 1 & BJG 1362 & \\ % 
1455 & 16.730 & 1.17 & 4.81 & 0.24 & 0.021 &  M & 2 & BJG 2505 & Noisy, PDM picks $\sim 9.6$\,d\\ % Pwrat 9.76d also
1456 & 16.730 & 1.13 & 13.31 & 1.58 & 0.026 &  M & 1 & BJG 4598 & \\ % Bang on MS!
1605 & 16.946 & 1.22 & 5.81 & 0.30 & 0.083 &  M & 1 & BJG 4556 & \\ % 
1711 & 17.091 & 1.22 & 10.22 & 0.84 & 0.063 &  M & 1 & BJG 3794 & \\ % 
1744 & 17.130 & 1.23 & 6.79 & 0.44 & 0.026 & -- & 2 &  & Noisy, PDM picks $\sim 13.3$\,d\\ % Noisy
2010 & 17.483 & 1.25 & 5.82 & 0.26 & 0.056 & -- & 1 &  & \\ % Below MS?
2071 & 17.568 & 1.35 & 7.30 & 0.57 & 0.088 &  M & 1 & BJG 2439 & \\ % 
2285 & 17.805 & 1.35 & 3.26 & 0.10 & 0.075 &  N & 1 & BJG 2458 & \\ % 
2346 & 17.852 & 1.46 & 9.68 & 0.72 & 0.073 &  M & 1 & BJG 771 & \\ % Above MS?
2632 & 18.161 & 1.55 & 1.67 & 0.03 & 0.117 &  M & 1 & BJG 3428 & \\ % 
\end{longtable}
\tablefoot{Mem = M, N, -- respectively indicate BJG05 members, non-members, and stars without BJG05 membership information.\\ 
Q = 1, 2 respectively indicate high quality and lower quality periods.}
\end{longtab}
\end{footnotesize}

\appendix

\newpage
\section{Online appendix}

% Figure A1-1 available electronically only
\onlfig{
\begin{figure*}%f3
\includegraphics[width=\hsize]{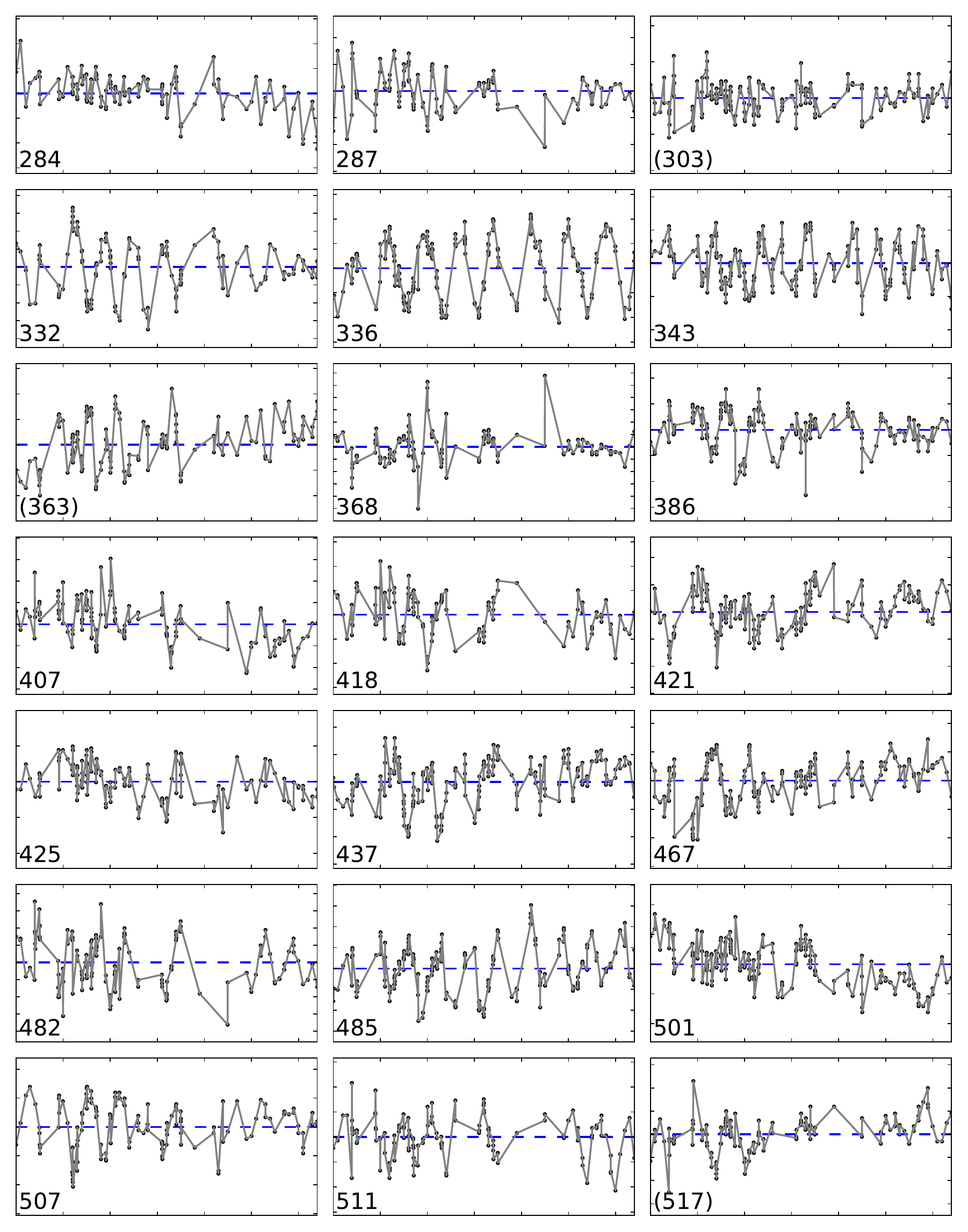}
\caption{Online lightcurves 1  (Q = 2 periods in parentheses, x-units = 10\,d, y-units = 0.01\,mag)}
\label{ltcrv21}
\end{figure*}
}

% Figure A1-2 available electronically only
\onlfig{
\begin{figure*}%f3
\includegraphics[width=\hsize]{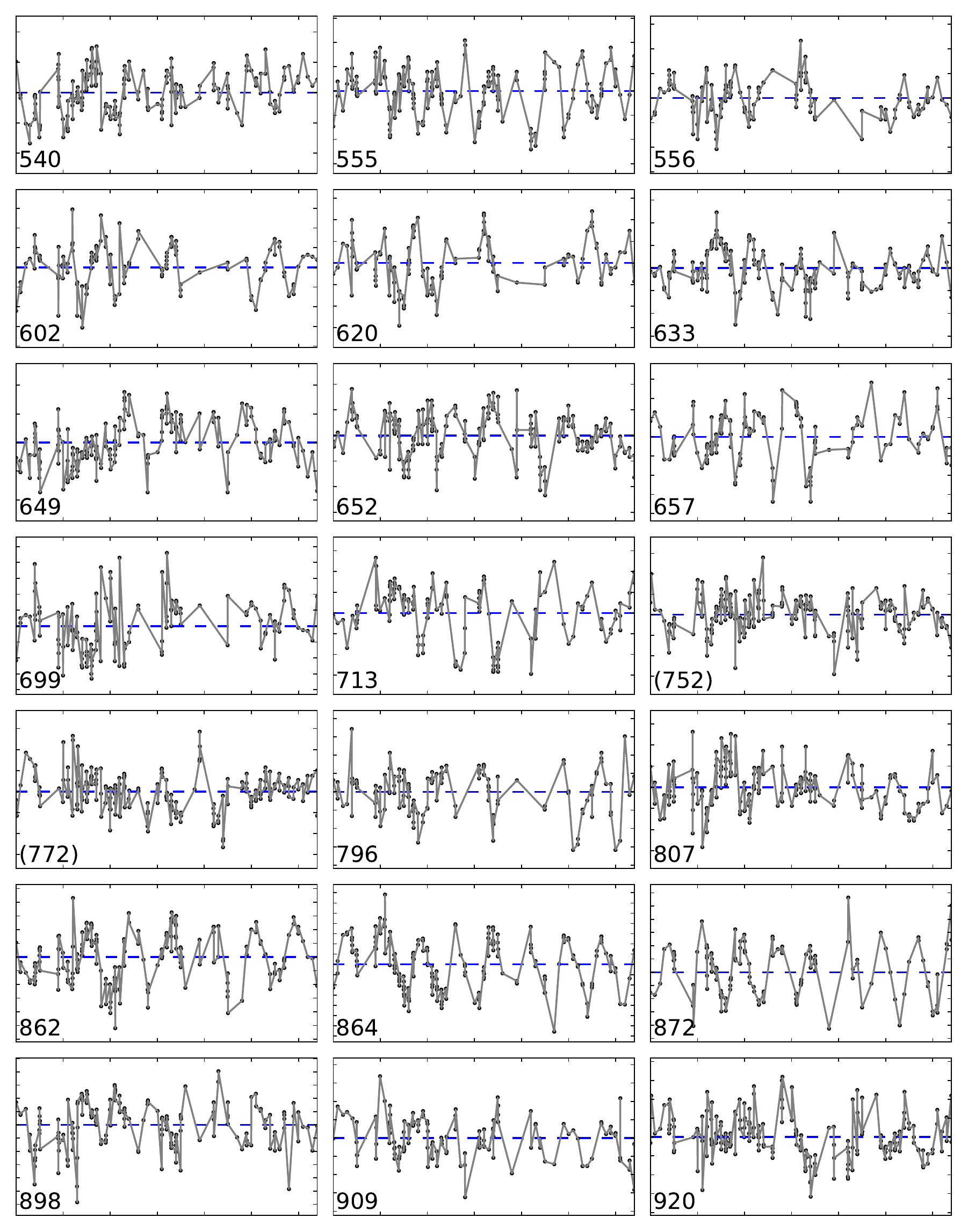}
\caption{Online lightcurves 2  (Q = 2 periods in parentheses, x-units = 10\,d, y-units = 0.01\,mag)}
\label{ltcrvs2}
\end{figure*}
}
% Figure A1-3 available electronically only
\onlfig{
\begin{figure*}%f3
\includegraphics[width=\hsize]{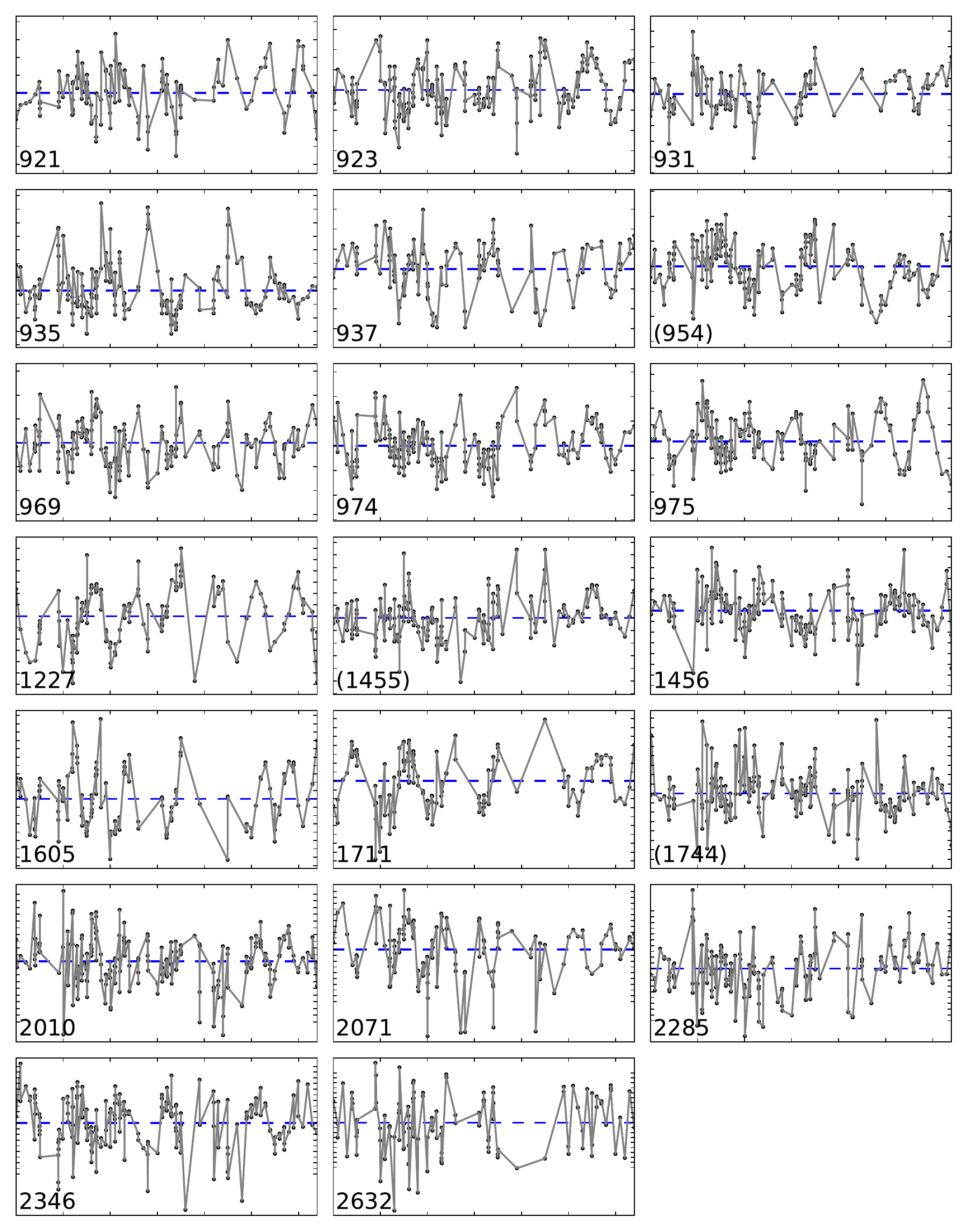}
\caption{Online lightcurves 3  (Q = 2 periods in parentheses, x-units = 10\,d, y-units = 0.01\,mag)}
\label{ltcrvs3}
\end{figure*}
}

% Figure A2-1 available electronically only
\onlfig{
\begin{figure*}%f3
\includegraphics[width=\hsize]{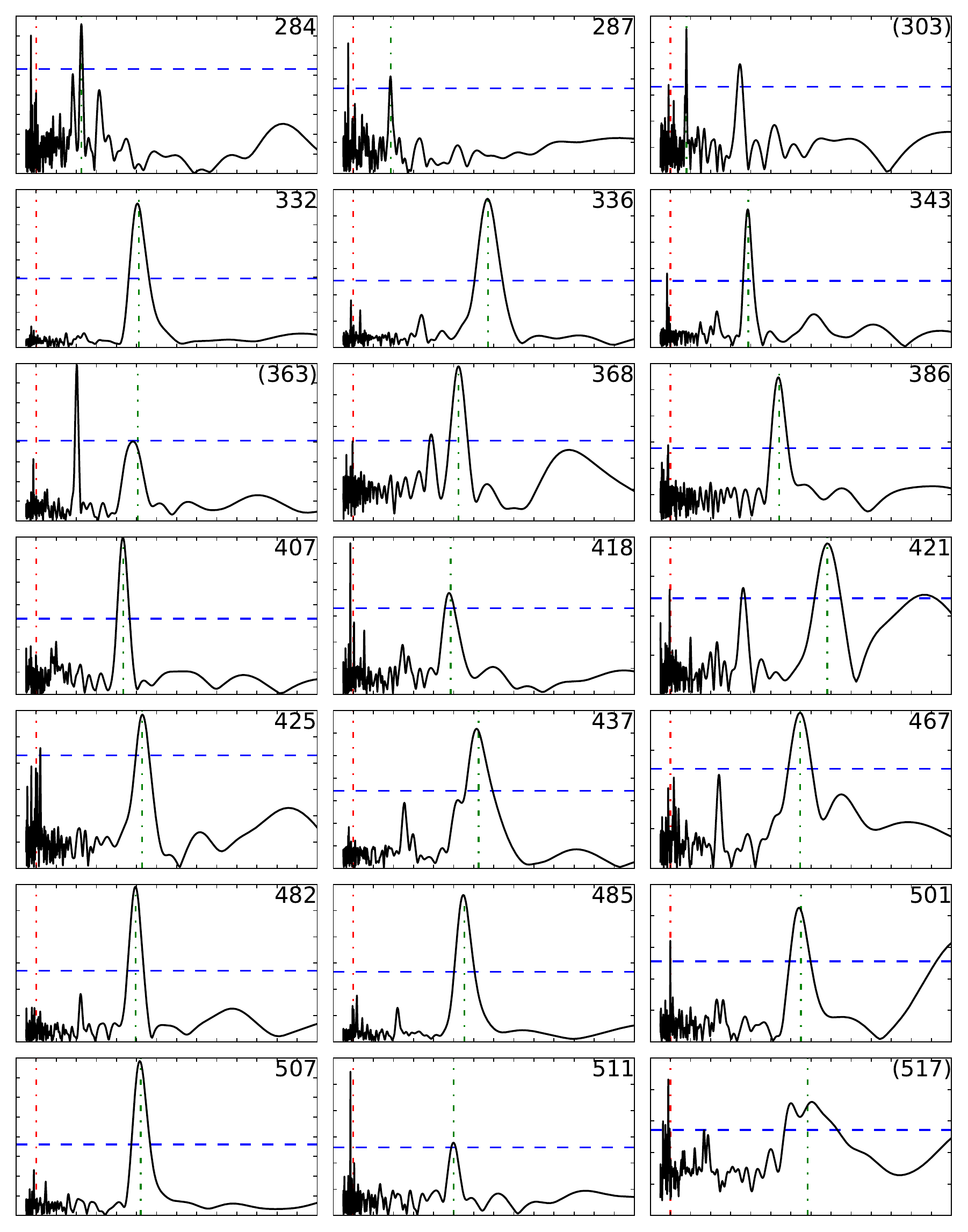}
\caption{Online spectra 1  (Q = 2 periods in parentheses, x-units = 1\,d, y-units arbitrary, selected rotation period marked with green line, 5\,$\sigma$ level marked with blue line}
\label{ltcrv21}
\end{figure*}
}

% Figure A2-2 available electronically only
\onlfig{
\begin{figure*}%f3
\includegraphics[width=\hsize]{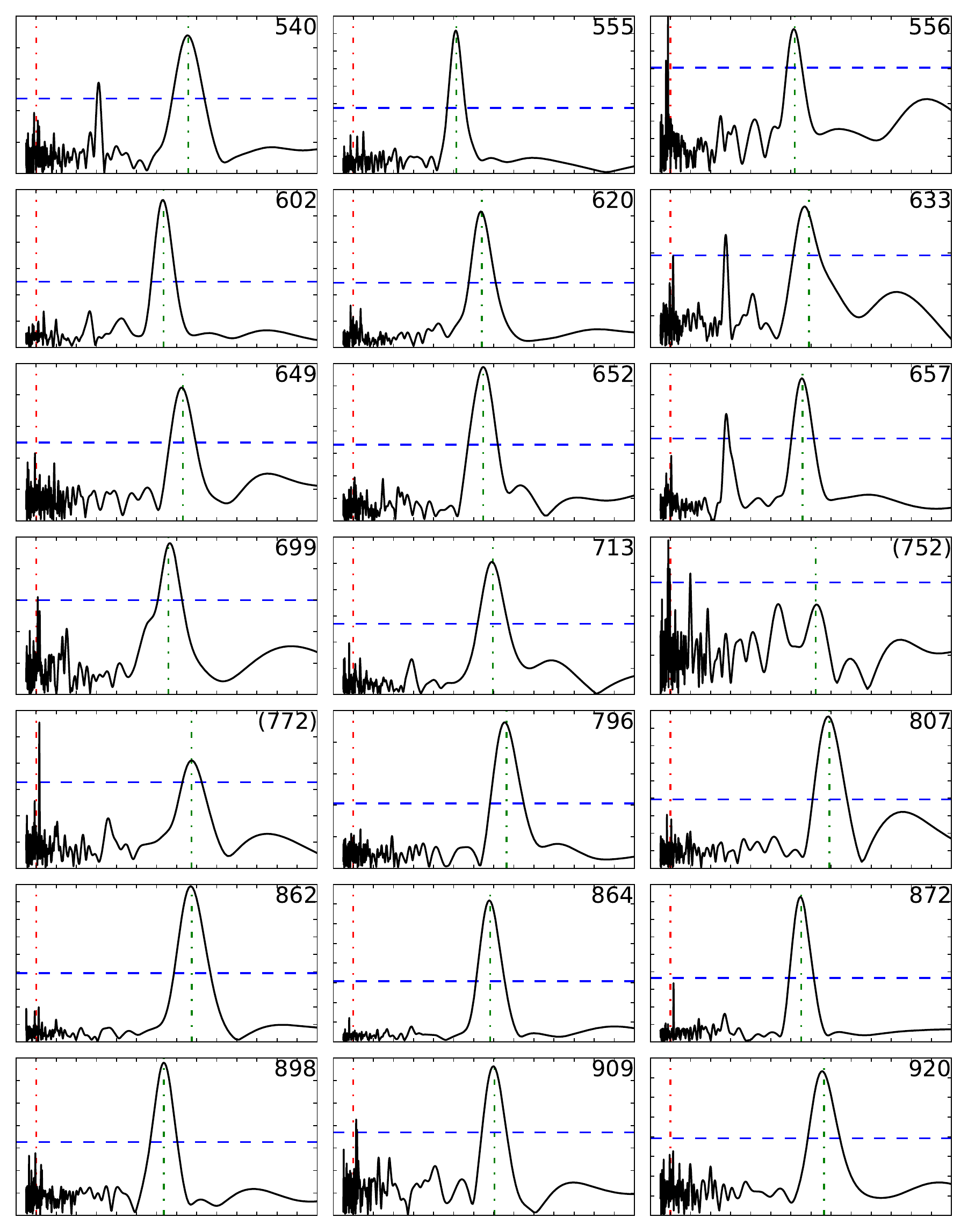}
\caption{Online spectra 2  (Q = 2 periods in parentheses, x-units = 1\,d, y-units arbitrary, selected rotation period marked with green line, 5\,$\sigma$ level marked with blue line)}
\label{ltcrvs2}
\end{figure*}
}
% Figure A2-3 available electronically only
\onlfig{
\begin{figure*}%f3
\includegraphics[width=\hsize]{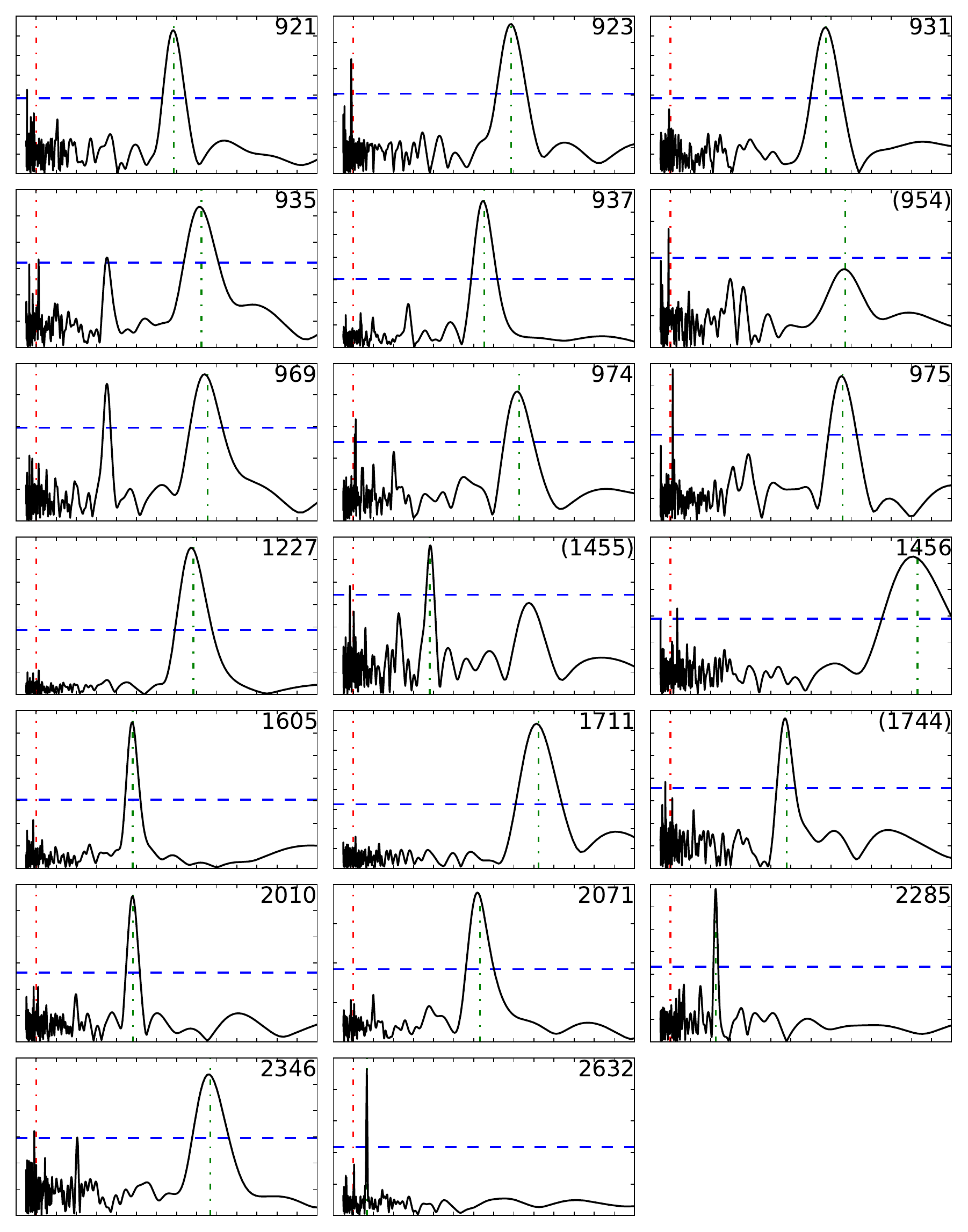}
\caption{Online spectra 3  (Q = 2 periods in parentheses, x-units = 1\,d, y-units arbitrary, selected rotation period marked with green line, 5\,$\sigma$ level marked with blue line)}
\label{ltcrvs3}
\end{figure*}
}

% Figure A3-1 available electronically only
\onlfig{
\begin{figure*}%f3
\includegraphics[width=\hsize]{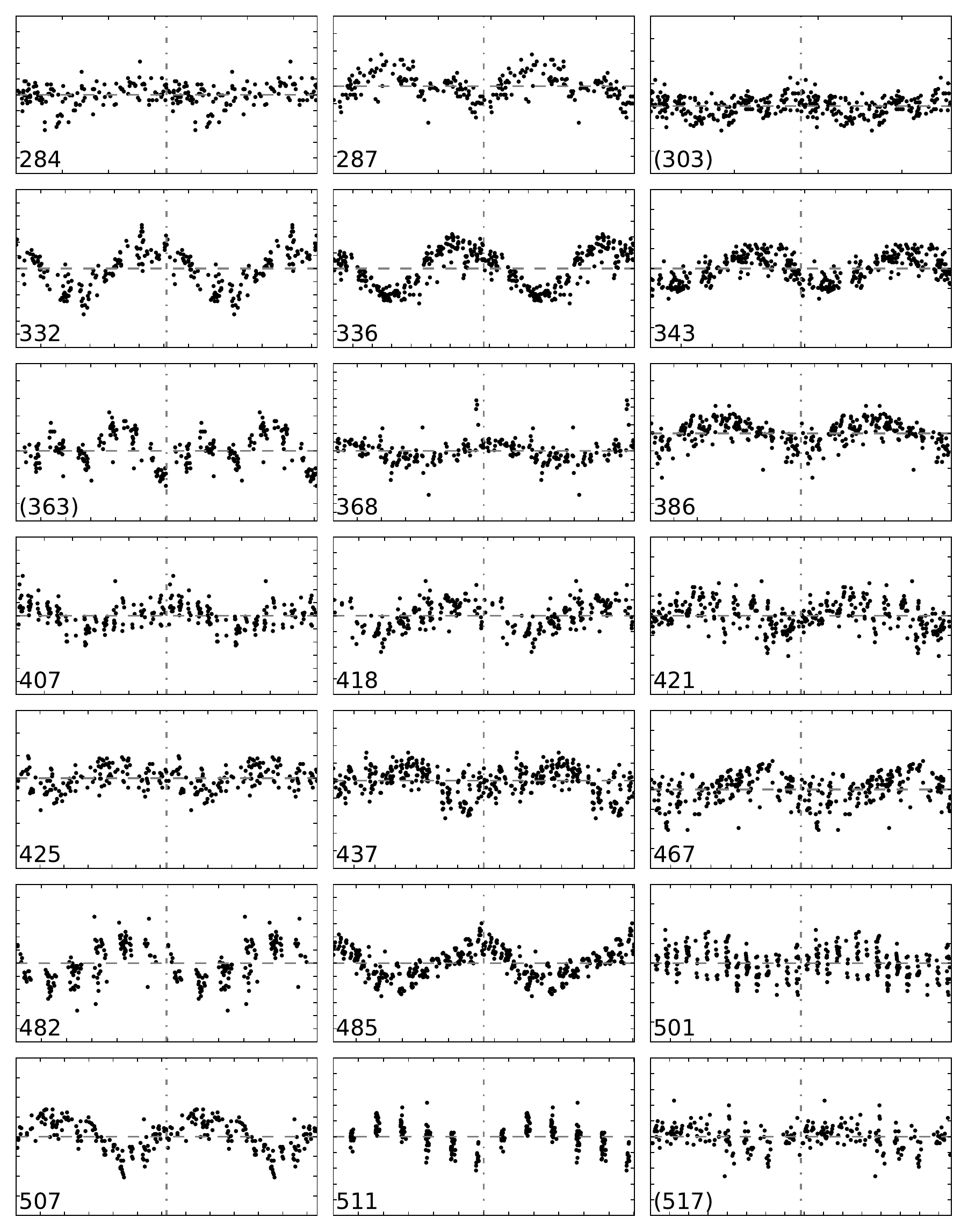}
\caption{Online phased curves 1  (Q = 2 periods in parentheses, x-units = 1\,d, y-units = 0.01\,mag), with two phases displayed for each star}
\label{phased1}
\end{figure*}
}
% Figure A3-2 available electronically only
\onlfig{
\begin{figure*}%f3
\includegraphics[width=\hsize]{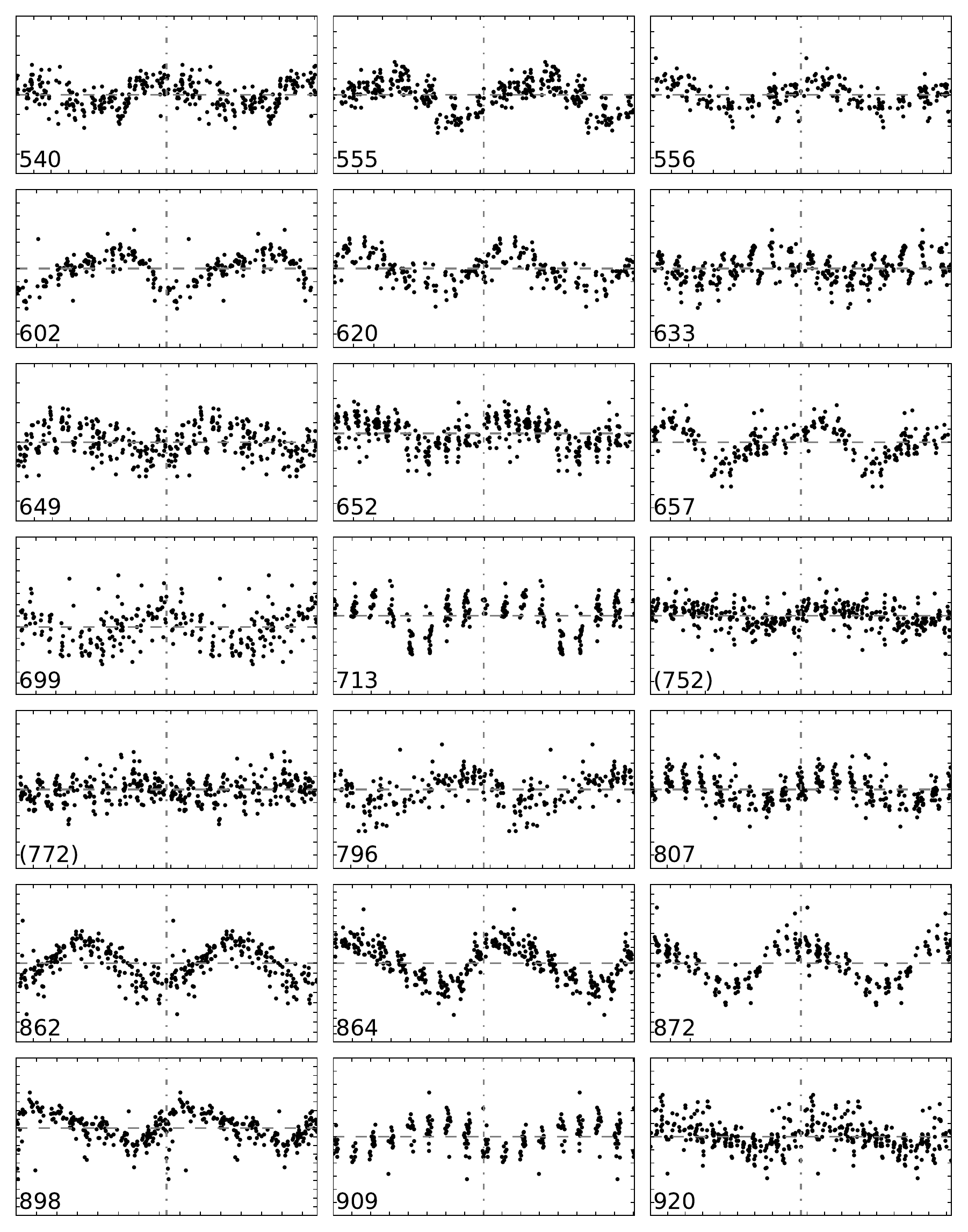}
\caption{Online phased curves 2  (Q = 2 periods in parentheses, x-units = 1\,d, y-units = 0.01\,mag), with two phases displayed for each star}
\label{phased2}
\end{figure*}
}
% Figure A3-3 available electronically only
\onlfig{
\begin{figure*}%f3
\includegraphics[width=\hsize]{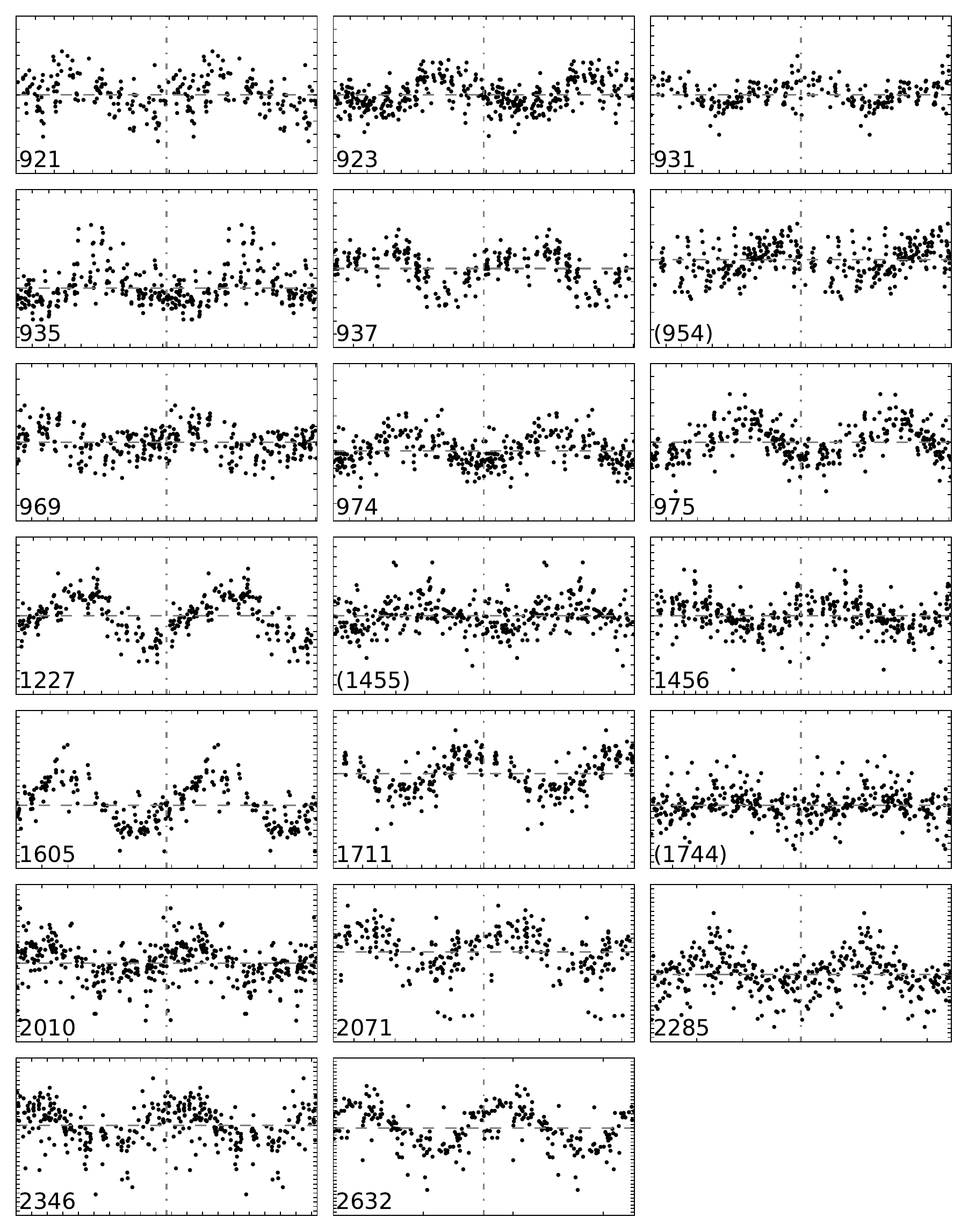}
\caption{Online phased curves 3  (Q = 2 periods in parentheses, x-units = 1\,d, y-units = 0.01\,mag), with two phases displayed for each star}
\label{phased3}
\end{figure*}
}
% Figure A2-4 available electronically only
%\onlfig{
%\begin{figure*}%f3
%\includegraphics[width=\hsize]{figs_submitted/phase4}
%\caption{Online phased curves 4}
%\label{phased4}
%\end{figure*}
%}

%__________________________________________________________________


\begin{thebibliography}{}

% FS: model atmospheres paper
\bibitem[]{} Allard, F., Homeier, D., \& Freytag, B.\ 2011 in 16th Cambridge Workshop on Cool Stars, Stellar Systems, and the Sun, ed. C. Johns-Krull, M. Browning, A. West, ASP, 448, 91 

\bibitem[]{}Angus, R., Aigrain, S., Foreman-Mackey, D., McQuillan, A., 2015, MNRAS, 450, 1787

\bibitem[yr]{code}Balaguer-Nunez, L., Jordi, C., \& Galadi-Enriquez, D., 2005 (BJG05), A\&A, 437, 457

\bibitem[]{}Baliunas, S. L., Sokoloff, D. \& Soon, W. H., 1996, ApJL, 457, 99

\bibitem[]{}Barnes, S. A., 2003, ApJ, 586, 464

\bibitem[]{}Barnes, S. A., 2007, ApJ, 669, 1167

\bibitem[]{}Barnes, S. A., 2010 (B10), ApJ, 722, 222

\bibitem[]{}Barnes, S. A. \& Kim, Y.-C., 2010, ApJ, 721, 675 

\bibitem[]{}Barnes, S. A. \& Sofia, S., 1996, ApJ, 462, 746 

\bibitem[]{}Barnes, S. A., Sofia, S., Stauffer, J. \& Prosser, C. P. 1999, ApJ, 516, 263 %IC2602

\bibitem[]{}Bertin, E. \& Arnouts, S., 1996, A\&A, 117, 393

\bibitem[]{}Bouvier, J., Cabrit, S., Fernandez, M., Martin, E.L., Matthews, J.M., 1993, A\&A, 272, 176

\bibitem[]{}Bouvier, J., Forestini, M. \& Allain, S., 1997, A\&A, 326, 1023 

\bibitem[]{}Brown, T. M. 2014, ApJ, 789, 101 %Metastable Dynamo

\bibitem[]{}Calabretta, M. R. \& Greisen, E. W., 2002, A\&A, 395, 1077

\bibitem[]{}Canterna, R., Perry, C. L., \& Crawford, D. L. 1979, PASP, 91, 263

\bibitem[]{}Cargile, P., James, D. J., Pepper, J. et al. 2014, ApJ, 782, 29 %(Blanco\,1)

\bibitem[]{}Chaboyer, B., Demarque, P., \& Pinsonneault, M. H., 1995, ApJ, 441, 876 

\bibitem[]{}Collier Cameron, A.,  Campbell, C. G., Quaintrell, H., 1995, A\&A, 298, 133


\bibitem[]{}Chromey, F. R. \& Hasselbacher, D. A., 1996, PASP, 108, 944

\bibitem[]{}Crawford, D. L. 1975, AJ, 80, 955

\bibitem[]{}Delorme, P., Cameron, A. C., Hebb, L., et al. 2011, MNRAS, 413, 2218 %Hyades

\bibitem[]{}Ebbighausen, E. G. 1939, ApJ, 90, 689

\bibitem[]{}Epstein, C. \& Pinsonneault, M., 2014, ApJ, 780, 159

\bibitem[]{}Fossati, L., Bagnulo, S., Landstreet, J., Wade, G., Kochukov, O., Monier, R., Weiss, W., \& Gebran M., 2008, A\&A, 483, 891

\bibitem[]{}Gallet, F. \& Bouvier, J., 2015, A\&A, 577, 98

\bibitem[]{}Granzer, T. 2004, AN, 325, 513

\bibitem[]{}Hartman, J. D., Bakos, G. A., Kovacs, G. \& Noyes, R. W. 2010, MNRAS, 408, 475 %Pleiades CPD paper

\bibitem[]{}Hartman, J.D., Gaudi, B.S., Pinsonneault, M.H. et al., 2009, ApJ, 691, 342 %M\,37 paper

% Fs: PHOENIX paper
\bibitem[]{}Hauschildt, P.~H., Allard, F., \& Baron, E. 1999, ApJ, 512, 377 

\bibitem[]{}Ianna, P. A., \& Schlemmer, D. M. 1993, AJ, 105, 209

\bibitem[]{}Irwin, J., Hodgkin, S., Aigrain, S. et al., 2007, MNRAS, 377, 741

\bibitem[]{}James, D. J., Barnes, S. A., Meibom, S. et al. 2010, A\&A, 515, 100

\bibitem[]{}Kovacs, G, Hartman, J. D., Bakos, G. A. et al. 2014, MNRAS, 442, 2081

\bibitem[]{}Kraft, R., 1970, in Spectroscopic Astrophysics, An Assessment of the Contributions of Otto Struve. Edited by G.H. Herbig. Berkeley: University of California Press, p.385

\bibitem[]{}Landolt, A. 2009, AJ, 137, 4186

\bibitem[]{}MacGregor, K. B. \& Brenner, M., 1991, ApJ, 376, 204 

\bibitem[]{}Mamajek, E. \& Hillenbrand, L. A., 2008, ApJ, 687, 1264

\bibitem[]{}Matt, S., Brun, A. S., Baraffe, I. et al. 2015, ApJL, 799, 23

\bibitem[]{}Meibom, S., Mathieu, R. D. \& Stassun, K. G. 2009, ApJ, 653, 621 %M35 CPD paper

\bibitem[]{}Meibom, S., Mathieu, R. D., Stassun, K. G., Liebesny, P. \& Saar, S. H. 2011a, ApJ, 733, 115 %M34 CPD paper

\bibitem[]{}Meibom, S., Barnes, S. A., Latham, D. W. et al. 2011b, ApJL, 733, 9 %NGC 6811

\bibitem[]{}Meibom, S., Barnes, S. A., Platais, I. et al., 2015, Nature, 517, 589 %NGC 6819

\bibitem[]{}Mink, D. J., 2002, in Astronomical Data Analysis Software and Systems XI, ASP Conference Proceedings, Vol. 281. Edited by David A. Bohlender, Daniel Durand, and Thomas H. Handley. ISBN: 1-58381-124-9. ISSN: 1080-7926. San Francisco: Astronomical Society of the Pacific, 2002, p. 169.

\bibitem[]{}Patten, B. M. \& Simon, T. 1996, ApJS, 106, 489 %IC2391

\bibitem[]{}Pesch, P. 1961, ApJ, 134, 602

\bibitem[]{}Perryman, M. A. C., Brown, A. G. A., Lebreton, Y. et al. 1998, A\&A, 331, 81

\bibitem[]{}Queloz, D., Allain, S., Mermilliod, J.-C., Bouvier, J., \& Mayor, M., 1998, A\&A, 335, 183

\bibitem[]{}Radick, R. R., Thompson, D. T., Lockwood, G. W., Duncan, D. K. \& Baggett, W. E., 1987, ApJ, 321, 459

\bibitem[]{}Rider, C. J., Tucker, D. L., Smith, J. A. et al. 2004, AJ, 127, 2210

\bibitem[]{}Roberts, D. H., Lehar, J. \& Dreher, J. W., 1987, 93, 968

\bibitem[]{}Sharma, S., Pandey, A. K., Ogura, K. et al. 2006, AJ, 132, 1669

\bibitem[]{}Sills, A., Pinsonneault, M. H., \& Terndrup, D. M., 2000, ApJ, 534, 335

\bibitem[]{}Soderblom, D.R., Stauffer, J.R., Hudon, J.D., \& Jones, B.F., 1993, ApJS, 85, 315

\bibitem[]{}Spada, F., Demarque, P., Kim, Y.-C. \& Sills, A. 2013, ApJ, 776, 87

\bibitem[]{}Stauffer, J.R. \& Hartmann, L.W., 1987, ApJ, 318, 337

\bibitem[]{}Stellingwerf, R. F. 1978, ApJ, 224, 953

\bibitem[]{}Strassmeier, K. G., Granzer, T., Weber, M. et al., 2004, AN, 325, 527

\bibitem[]{}Strassmeier, K. G., Granzer, T., \& Weber, M., 2010, Adv. in Astr., 2010, p.19

\bibitem[]{}Strassmeier, K. G., Weingrill, J., Granzer, T. et al. 2015, A\&A, 580, 66 %IC\,4756

\bibitem[]{}Taylor, B. J. 2006, AJ, 132, 2453

\bibitem[]{}van Leeuwen, F. \& Alphenaar, P., 1982, Messenger, 28, 15

\bibitem[]{}van Leeuwen, F., Alphenaar, P. \& Meys, J.J.M., 1987, A\&AS, 67, 483

\bibitem[]{}Wilson, O. C., 1978, ApJ, 226, 379

\bibitem[]{}Wu, Z. Y., Tian, K. P., Balaguer-Nunez, L. et al. 2002, A\&A, 381, 464

\bibitem[]{}Wu, Z.-Y., Zhou, X., Ma, J., Jiang, Z.-J., \& Chen, J.-S. 2006, PASP, 118, 1104

\bibitem[]{}Yi, S., Demarque, P., Kim, Y.-C. et al., 2001, ApJS, 136, 417 %Y^2 isochrones

\bibitem[]{}Zechmeister, M. \& Kurster, M., 2009, A\&A, 496, 577 %Generalized Lomb-Scargle

\end{thebibliography}
\end{document}